\documentclass[twocolumn,superscriptaddress,amsfonts,nopacs,amsmath,amssymb]{revtex4-1}
\usepackage[T1]{fontenc}
\usepackage[latin9]{inputenc}
\usepackage{bm}
\usepackage{amsmath}
\usepackage{amssymb}
\usepackage{graphicx}

\makeatletter

\providecommand{\tabularnewline}{\\}

\pdfoutput=1
\usepackage{ucs}
\usepackage{graphics}
\usepackage{dcolumn}
\usepackage{epsfig}
\usepackage{bm}
\usepackage{braket}

\makeatother

\begin{document}
\title{A minimal tight-binding model for\\
 the quasi-one-dimensional superconductor K$_{2}$Cr$_{3}$As$_{3}$}
\author{Giuseppe Cuono}
\affiliation{Dipartimento di Fisica ``E.R. Caianiello'', Università degli Studi
di Salerno, I-84084 Fisciano (Sa), Italy}
\author{Carmine Autieri}
\affiliation{International Research Centre MagTop, Institute of Physics, Polish
Academy of Sciences, PL-02668 Warsaw, Poland}
\author{Filomena Forte}
\affiliation{Dipartimento di Fisica ``E.R. Caianiello'', Università degli Studi
di Salerno, I-84084 Fisciano (Sa), Italy}
\affiliation{CNR-SPIN, c/o Università degli Studi di Salerno, I-84084 Fisciano
(Sa), Italy}
\author{Maria Teresa Mercaldo}
\affiliation{Dipartimento di Fisica ``E.R. Caianiello'', Università degli Studi
di Salerno, I-84084 Fisciano (Sa), Italy}
\author{Alfonso Romano}
\affiliation{Dipartimento di Fisica ``E.R. Caianiello'', Università degli Studi
di Salerno, I-84084 Fisciano (Sa), Italy}
\affiliation{CNR-SPIN, c/o Università degli Studi di Salerno, I-84084 Fisciano
(Sa), Italy}
\author{Adolfo Avella}
\affiliation{Dipartimento di Fisica ``E.R. Caianiello'', Università degli Studi
di Salerno, I-84084 Fisciano (Sa), Italy}
\affiliation{CNR-SPIN, c/o Università degli Studi di Salerno, I-84084 Fisciano
(Sa), Italy}
\affiliation{Unità CNISM di Salerno, Università degli Studi di Salerno, I-84084
Fisciano (Sa), Italy}
\author{Canio Noce}
\affiliation{Dipartimento di Fisica ``E.R. Caianiello'', Università degli Studi
di Salerno, I-84084 Fisciano (Sa), Italy}
\affiliation{CNR-SPIN, c/o Università degli Studi di Salerno, I-84084 Fisciano
(Sa), Italy}
\date{\today}
\begin{abstract}
We present a systematic derivation of a minimal five-band tight-binding
model for the description of the electronic structure of the recently
discovered quasi-one-dimensional superconductor K$_{2}$Cr$_{3}$As$_{3}$.
Taking as a reference the density-functional theory (DFT) calculation,
we use the outcome of a Löwdin procedure to refine a Wannier projection
and fully exploit the predominant weight at the Fermi level of the
states having the same symmetry of the crystal structure. Such states
are described in terms of five quasi-atomic $d$ orbitals: four planar
orbitals, two $d_{xy}$ and two $d_{x^{2}-y^{2}}$, and a single out-of-plane
one, $d_{z^{2}}$. We show that this minimal model reproduces with
great accuracy the DFT band structure in a broad energy window around
the Fermi energy. Moreover, we derive an explicit simplified analytical
expression of such model, which includes three nearest-neighbor hopping
terms along the $z$ direction and one nearest-neighbor term within
the $xy$ plane. This model captures very efficiently the energy spectrum
of the system and, consequently, can be used to study transport properties,
superconductivity and dynamical effects in this novel class of superconductors.
\end{abstract}
\maketitle

\section{Introduction}

The study of the interplay between superconductivity and magnetism
has recently brought to the discovery of superconductivity in chromium-based
compounds~\citep{Chen19}. The first example of such systems was
CrAs, where the superconducting transition takes place at $T_{c}=2\,$K
as a result of the suppression of the antiferromagnetic transition
upon applying high pressure~\citep{Wu10,Wu14,Kotegawa14,Nigro18}.
This experimental finding inspired the search for superconductivity
in other Cr-based materials, which led to the discovery of ambient-pressure
superconductivity at 6.1 K in a Cr-based arsenide, K$_{2}$Cr$_{3}$As$_{3}$,
and, subsequently, in a whole class represented by the family A$_{2}$Cr$_{3}$As$_{3}$,
with A being K~\citep{Bao15}, Rb~\citep{Tang15}, Cs~\citep{Ztang15}
or Na~\citep{Mu18}. Remarkably, these Cr-based superconductors have
a quasi-one-dimensional (Q1D) crystal structure that consists of {[}(Cr$_{3}$As$_{3}$)$^{2-}${]}$_{\infty}$
double-walled nanotubes in which chromium atoms form the inner wall
and arsenic atoms the outer one. These nanotubes are in turn separated
by columns of A$^{+}$ ions~\citep{Bao15,Tang15,Ztang15}.

These novel superconductors display intriguing physical properties,
both in the normal ~\citep{Bao15,Kong15} and in the superconducting
phase~\citep{Bao15,Zhi15,Adroja15,Pang15}, which are under intense
investigations especially to clarify the role played by the reduced
dimensionality and by the electronic correlations ~\citep{Bao15}.
This is of course an important issue since the latter are features
which considerably affect the properties of a large fraction of the
unconventional superconductors discovered so far.

\begin{figure}[t]
\centering \includegraphics[width=8cm]{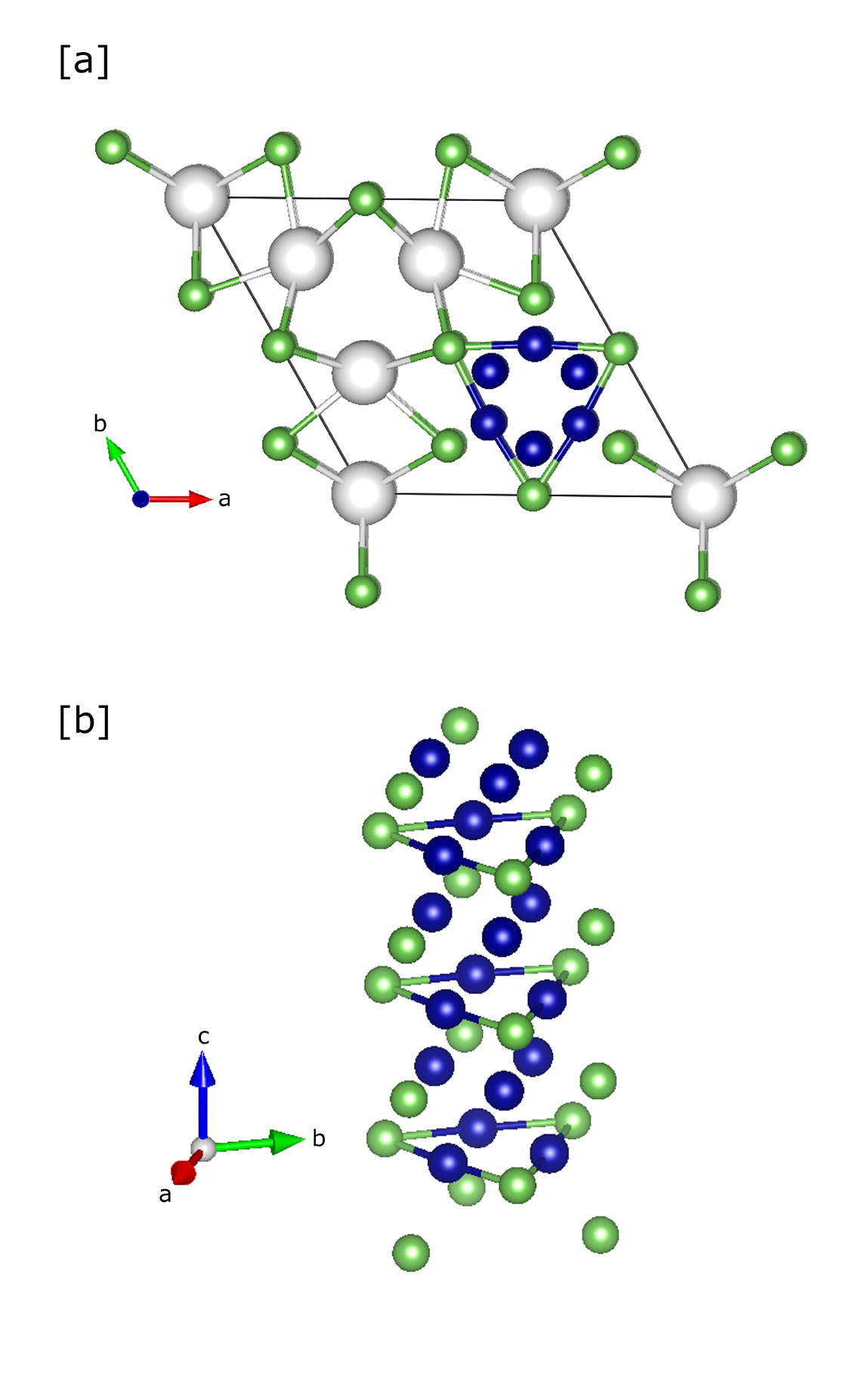} \caption{Crystal structure of K$_{2}$Cr$_{3}$As$_{3}$. The colours blue,
green and white denote Cr, As and K atoms, respectively. (a) Top view
of the primitive cell. (b) The double-walled nanotube formed by Cr
and As atoms.}
\label{fig:K2unitcell} 
\end{figure}

The crystal structure of K$_{2}$Cr$_{3}$As$_{3}$ is the hexagonal
one reported in Fig.~\ref{fig:K2unitcell}, with $a=9.9832$ Å and
$c=4.2304$ Å. The crystal structure exhibits two planes orthogonal
to the $c$ axis with slightly different stoichiometry, i.e. a plane
with KCr$_{3}$As$_{3}$ and a plane with K$_{3}$Cr$_{3}$As$_{3}$
stoichiometry~\citep{Bao15}. The resistivity of K$_{2}$Cr$_{3}$As$_{3}$
shows a linear temperature dependence in a broad temperature range,
which suggests a non-Fermi-liquid normal state, possibly related to
a quantum criticality and/or a realization of a Luttinger liquid~\citep{Bao15}.
This occurrence has not been confirmed by Kong \textit{et al.}~\citep{Kong15},
who rather report a $T^{3}$ dependence of the resistivity from 10
to 40 K. However, it should be noted that while this result refers
to measurements performed on single crystals, those reported in Ref.~\onlinecite{Bao15}
have been obtained on polycrystalline samples.

The superconductivity in K$_{2}$Cr$_{3}$As$_{3}$ shows various
features pointing towards an unconventional nature. An anisotropic
upper critical field is reported, with different amplitudes between
the cases of field applied parallel and perpendicular to the rodlike
crystals~\citep{Kong15}. NMR measurements of the nuclear spin-lattice
relaxation rate $1/T_{1}$ show a strong enhancement in the Cr nanotubes
of the spin fluctuations above $T_{c}$, with the power-law temperature
dependence $1/T_{1}T\sim T^{-\gamma}$ ($\gamma\simeq0.25$) being
consistent with a Tomonaga-Luttinger liquid~\citep{Zhi15}. In addition,
the absence of the Hebel-Slichter coherence peak in $1/T_{1}$ below
$T_{c}$ provides further evidence that the superconducting phase
is unconventional~\citep{Zhi15}. The same kind of indication comes
from muon-spin rotation measurements~\citep{Adroja15}, which provide
evidence of a possible $d$-wave superconducting pairing, as well
as from measurements of the temperature dependence of the penetration
depth $\Delta\lambda=\lambda(T)-\lambda(0)$~\citep{Pang15}. For
the latter, a linear behavior is observed for $T\ll T_{c}$, instead
of the exponential behavior of conventional superconductors, indicating
the presence of line nodes in the superconducting gap and thus supporting
the hypothesis of an unconventional nature of the superconducting
phase~\citep{Pang15}.

The unusual metallic state stimulated several studies with the aim
of attaining the best description of the system. ARPES studies~\citep{Watson17}
of single crystals reveal two Q1D Fermi surface sheets with linear
dispersions, without indication of any three dimensional 3D Fermi
surface, as instead predicted by Density Functional Theory (DFT) calculations~\citep{Jiang15}.
The overall bandwidth of the Cr 3$d$ bands and the Fermi velocities
are comparable to DFT results, indicating that the correlated Fermi
liquid picture is not appropriate for K$_{2}$Cr$_{3}$As$_{3}$.
Furthermore, the spectral weight of the Q1D bands decreases near the
Fermi level according to a linear power law, in an energy range of
200 meV. This result has been interpreted as an issue supporting a
Tomonaga-Luttinger liquid behavior.

On the other hand, measurements and modelling of K$_{2}$Cr$_{3}$As$_{3}$
spin wave excitations show that inter-tube $J$ terms are necessary
to reproduce the experimental data~\citep{Taddei17}. Furthermore,
using DFT, it has been found that in-plane structural distortions,
driven by unstable optical phonon modes, play an important role to
control the subtle interplay between the structural properties, the
electron-phonon and the magnetic interactions~\citep{Taddei18}.
These results point out the importance of both the intra- and the
inter-tube dynamics, as well as the relevance of the electron-phonon
and the magnetic interactions.

From theory perspective, the electronic structure of K$_{2}$Cr$_{3}$As$_{3}$
has been examined through density functional theory calculations~\citep{Jiang15}.
In contrast with other Q1D superconductors, K$_{2}$Cr$_{3}$As$_{3}$
exhibits a relatively complex electronic structure, where the Cr-3$d$
orbitals, specifically the $d_{z^{2}}$ , $d_{xy}$ and $d_{x^{2}-y^{2}}$
ones, dominate the electronic states near the Fermi energy~\citep{Jiang15}.
Several related calculations have been also developed~\citep{XWu15,Zhang16,Zhong15,Zhou17}
which establish a basis for theoretical models. In particular, it
has been shown~\citep{XWu15}, by using a three-band model built
from the above mentioned $3d$ orbitals, that a triplet $p_{z}$-wave
pairing driven by ferromagnetic fluctuations is the leading pairing
symmetry for physically realistic parameters. This result, holding
in both weak and strong coupling limits, has been confirmed in a subsequent
paper where a more accurate six-band model was used~\citep{Zhang16}.
Another theoretical work focuses on the study of a twisted Hubbard
tube modelling the {[}(Cr$_{3}$As$_{3}$)$^{2-}${]}$_{\infty}$
structure~\citep{Zhong15}. Here, a three-channel effective Hamiltonian
describing a Tomonaga-Luttinger liquid is derived and it is shown,
within this scenario, that the system tends to exhibit triplet superconducting
instabilities within a reasonable range of the interaction parameters.
Finally, the superconducting phase has also been investigated by means
of an extended Hubbard model with three molecular orbitals in each
unit cell~\citep{Zhou17}; as in the previously mentioned approaches,
it is found that the dominant pairing channel is always a spin-triplet
one, both for small and large $U$.

In this paper, we present the construction of a minimal tight-binding
(TB) model hamiltonian, which reproduces with high accuracy the band
structure of K$_{2}$Cr$_{3}$As$_{3}$ around the Fermi level, as
obtained via first-principle calculations. We demonstrate that such
description can be derived starting from a minimal set of five quasi-atomic
orbitals with mainly $d$ character, namely four planar orbitals ($d_{xy}$
and $d_{x^{2}-y^{2}}$ for each of the two planes KCr$_{3}$As$_{3}$
and K$_{3}$Cr$_{3}$As$_{3}$) and a single out-of-plane one ($d_{z^{2}}$).
Moreover, we derive an explicit simplified analytical expression of
our five-band TB model, which includes three nearest-neighbor (NN)
hopping terms along the $z$ direction and one NN within the $xy$
plane.

Such TB representation of the K$_{2}$Cr$_{3}$As$_{3}$ band structure
is obtained within a three-stage approach consisting of the following
steps: (i) Guided by first-principles DFT calculations, we first construct
the model based on the atomic Cr and As orbitals and use it to investigate
the orbital character and the symmetry of the bands which dominate
in a certain energy window around the Fermi level; (ii) we use the
Löwdin procedure to downfold the original full hamiltonian into a
much smaller space spanned by a set of atomic Cr and As orbitals which
are symmetric with respect to the basal plane; iii) the knowledge
of the results of steps i) and ii) allows to formulate, within a Wannier
projection, a TB description based on five atomic-like orbitals of
mainly $d$ character, which are spatially localized around virtual
lattice sites located at the center of the Cr-triangles of the K$_{3}$Cr$_{3}$As$_{3}$
planes, stacked along the chain direction.

We point out that this minimal model well reproduces all the details
of the low-energy band structures in a broad energy region around
the Fermi level, which makes it a remarkable improvement with respect
to previous three-band models. Moreover, we expect that, by solving
such minimal model and its extensions using suitable approximations,
one may obtain information about the superconducting pairing mechanism
especially for the pairing symmetry, starting from a more complete
band structure.

The paper is organized as follows. In the next Section, we present
the details of the DFT calculations as well as their extension taking
into account the effect of the local Hubbard interaction. In Sec.~III,
we present the TB description in terms of atomic orbitals, which serves
as a starting point for the study of the total and the local density
of states, together with the characterization of the Cr and As orbital
components of the electronic bands. In Sec.~IV, we present the results
of the Löwdin down-folding procedure giving the projection of the
total Hamiltonian on the subspace of the orbitals that dominate at
the Fermi level. From the results of this last section, we are able
to identify the relevant orbitals at the Fermi level, so that we formulate
a minimal five-band effective tight-binding model, as described in
Section~V, whereas last Section contains our conclusions.

\section{Density functional calculations}

In this section, we present the first-principle calculations which
supply a basis for constructing the TB modeling of the K$_{2}$Cr$_{3}$As$_{3}$
band structure that will be described in the following sections. The
real space Hamiltonian matrix elements have been set according to
the outcome of DFT calculations~\citep{CAutieri17}, performed by
using the VASP package~\citep{vasp}. In such an approach, the core
and the valence electrons have been treated within the projector augmented
wave method~\citep{vasp1} and with a cutoff of 500~eV for the plane
wave basis. All the calculations have been performed using a 4$\times$4$\times$10
$k$-point grid. For the treatment of the exchange correlation, the
local density approximation and the Perdew-Zunger~\citep{Perdew}
parametrization of the Ceperley-Alder~\citep{Ceperley} data have
been considered. After obtaining the Bloch wave functions, the maximally
localized Wannier functions~\citep{Marzari97,Souza01} are constructed
using the WANNIER90 code~\citep{Mostofi08}. To extract the Cr 3$d$
and As 4$p$ electronic bands, the Slater-Koster interpolation scheme
has been used, in order to determine the real-space Hamiltonian matrix
elements~\citep{Mostofi08}.

\begin{figure}
\centering \includegraphics[width=8cm]{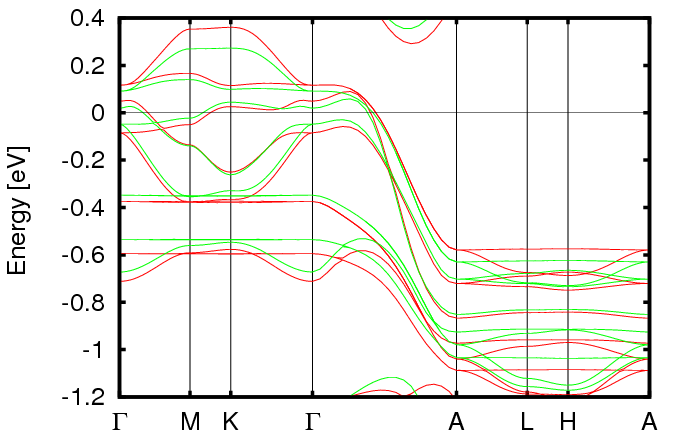} \caption{LDA+U band structure near the Fermi level for $U=0$ (green lines)
and $U=4$ eV (red lines).}
\label{fig:bdft} 
\end{figure}

The role of the electronic correlations on the energy spectrum of
K$_{2}$Cr$_{3}$As$_{3}$ has also been explored. To this purpose,
we have performed first-principle calculations taking into account
the effect of the local Hubbard interactions, assumed to be non-vanishing
on all the Cr $d$ orbitals, and efficiently parametrized by a finite
number of Slater integrals. We follow the convention~\citep{Anisimov93}
of identifying $U$ with the Slater integral $F^{0}$ and the Hund
coupling with $F^{2}$ and $F^{4}$. The direct calculation gives
for the intra-$t_{2g}$ orbital Hund interaction the value $J_{H}$
= 0.15 $U$~\citep{Vaugier12}. Considering that previous studies~\citep{Jiang15}
indicate that the system is weakly correlated, we have assumed for
$U$ values ranging in a interval going from 0 to 4 eV. The band structure
obtained in the two limiting cases of $U=0$ and $U=4$ eV is reported
in Fig.~\ref{fig:bdft}. As far as the $U=0$ case is concerned,
the results are in agreement with the literature~\citep{Jiang15}.
The character of the bands at low energies is mainly due to the $d$
states of Cr atoms, whereas the $p$ states of the As atoms are located
few electronvolts above and below the Fermi level, as also found for
CrAs~\citep{Autieri17,Autieri18,CAutieri17}.

We can see that, apart from a slight increase of the energy bandwidth
corresponding to the chromium states being pushed away from the Fermi
level, electronic correlations on chromium orbitals barely affect
the energy spectrum. In particular, the band energy separation occurring
when the Coulomb repulsion is turned on is one order of magnitude
lower than the value of $U$ considered in the calculation. This result
thus seems to confirm that K$_{2}$Cr$_{3}$As$_{3}$ is in a moderate-coupling
regime, characterized by a robust metallic phase, expected to remain
stable also under the influence of pressure, strain or doping. As
also found for CrAs~\citep{CAutieri17}, the nonmagnetic and the
antiferromagnetic phases turn out to be very close in energy. In the
case of K$_{2}$Cr$_{3}$As$_{3}$, the triangular geometry tends
to frustrate antiferromagnetism, so that the nonmagnetic phase is
the most stable one.

\section{Tight binding approximation and orbital characterization of the band
structure}

In this section, we derive a TB model obtained from a basis set of
localized atomic orbitals at each site of the crystal structure. Our
starting point will be the most reliable TB description that is capable
to reproduce the LDA band structure close to the Fermi level. This
description will then be used to examine in detail the orbital character
of the energy bands, which will be resolved both with respect to the
energy itself and to the $k$ vector in the main high-symmetry points
of the Brillouin zone.

We construct the TB model by considering all the atomic orbitals that
participate to conduction, namely Cr 3$d$ and As 4$p$ orbitals.
The basis in the Hilbert space is given by the vector 
\begin{equation}
\phi_{i}^{\dagger}=(d_{i,xy}^{\dagger},d_{i,x^{2}-y^{2}}^{\dagger},d_{i,z^{2}}^{\dagger},p_{i,x}^{\dagger},p_{i,y}^{\dagger},d_{i,yz}^{\dagger},d_{i,xz}^{\dagger},p_{i,z}^{\dagger}),\label{eqn:vectors}
\end{equation}
where $i$ is the lattice index and the orbitals are ordered having
first those symmetric with respect to the basal plane and then the
antisymmetric ones.

The tight-binding Hamiltonian is defined as 
\begin{equation}
H=\sum_{i}\phi_{i}^{\dagger}\hat{\varepsilon}_{i}\phi_{i}+\sum_{i,j}\phi_{i}^{\dagger}\hat{t}_{i,j}\phi_{j}\label{eqn:tightbinding}
\end{equation}

\noindent where $i$ and $j$ denote the positions of Cr or As atoms
in the crystal, and $\hat{\varepsilon}_{i}$ and $\hat{t}_{i,j}$
are matrices whose elements have indices associated with the different
orbitals involved. The first term of the Hamiltonian takes into account
the on-site energies, with $\left(\hat{\varepsilon}_{i}\right)^{\alpha\beta}=\varepsilon_{i}^{\alpha}\delta_{\alpha,\beta}$,
while the second term describes hopping processes between distinct
orbitals, with amplitudes given by the matrix elements $t_{ij}^{\alpha\beta}$.
The latter are given by the expectation values of the residual lattice
potential $V({\bf r})$ on the complete orthogonal set of the Wannier
functions $\phi_{\alpha}({\bf r}-{\bf R}_{i})$: 
\begin{equation}
t_{ij}^{\alpha\beta}=\bigl\langle\phi_{\alpha}({\bf r}-{\bf R}_{i})|V({\bf r})|\phi_{\beta}({\bf r}-{\bf R}_{j})\bigr\rangle\;.\label{eqn:hoppingparameters}
\end{equation}
Their values are obtained according to the procedure described in
the previous Section. The Hamiltonian in Eq.~(\ref{eqn:tightbinding})
is a $48\times48$ matrix because the primitive cell contains six
Cr and six As atoms and we have to consider five $d$ orbitals for
each Cr atom and three $p$ orbitals for each As atom.

Recently, we have carried out a detailed TB analysis in order to address
the nature of the electronic bands provided by ab-initio calculations,
in particular with respect to its supposed one-dimensionality~\citep{Cuono18}.
Such study revealed that considering only the hoppings between the
orbitals of the atoms that lie within a single sub-nanotube fails
completely to describe the in-plane band structure, not even allowing
the correct description of the band structure along the $z$ axis.
Such result is also in agreement with previous DFT calculations showing
that, in contrast with other quasi-1D superconductors, K$_{2}$Cr$_{3}$As$_{3}$
exhibits a relatively complex electronic structure and the Fermi surface
contains both 1D and 3D components~\citep{Jiang15}.

In order to understand how the band structure is affected by the number
and the position of the lattice cells involved in the hopping processes,
here we have carried out a more accurate optimization of the TB hamiltonian,
as explained in detail in Appendix A. Such procedure starts by first
considering the hopping processes within the quasi one-dimensional
{[}(Cr$_{3}$As$_{3}$)$^{2-}${]}$^{\infty}$ double-walled nanotubes
only, and then including step by step inter-tube and longer-range
intra-tube processes. Our analysis suggests that the LDA results arise
from a delicate combination of several very small contributions, which
are crucial in order to faithfully determine the dispersion of the
bands that cut the Fermi level perpendicular to the chain direction
($\Gamma MK\Gamma$ path). We thus conclude that, in order to obtain
a faithful representation of the electronic ab-initio band structure,
it is necessary to take into account all hopping processes up to the
fifth-neighbor cells along the $z$-axis, together with the in-plane
hoppings up to the second-neighbor cells. Diagonalizing the Hamiltonian
in Eq.~\ref{eqn:tightbinding} and retaining such hopping terms,
we obtain an energy spectrum that perfectly matches the one for $U=0$
in Fig.~\ref{fig:bdft} (see Fig.~\ref{fig:longrangezoomz}), as
evaluated along the high symmetry path of the hexagonal Brillouin
zone considered in Ref.~\onlinecite{Setyawan10}. Accordingly, the
results presented in this section have been obtained within this framework.
However, although the agreement is extremely satisfactory, we cannot
consider successfully concluded our quest for a minimal model because
of the need for so many hopping parameters. Therefore, in order to
gain sufficient insight in the behavior of the system and design an
efficient reduction procedure leading us to a real minimal model,
we proceed with the analysis of the partial density of states and
of the orbital character of the bands.

\begin{figure}
\centering \includegraphics[width=7cm]{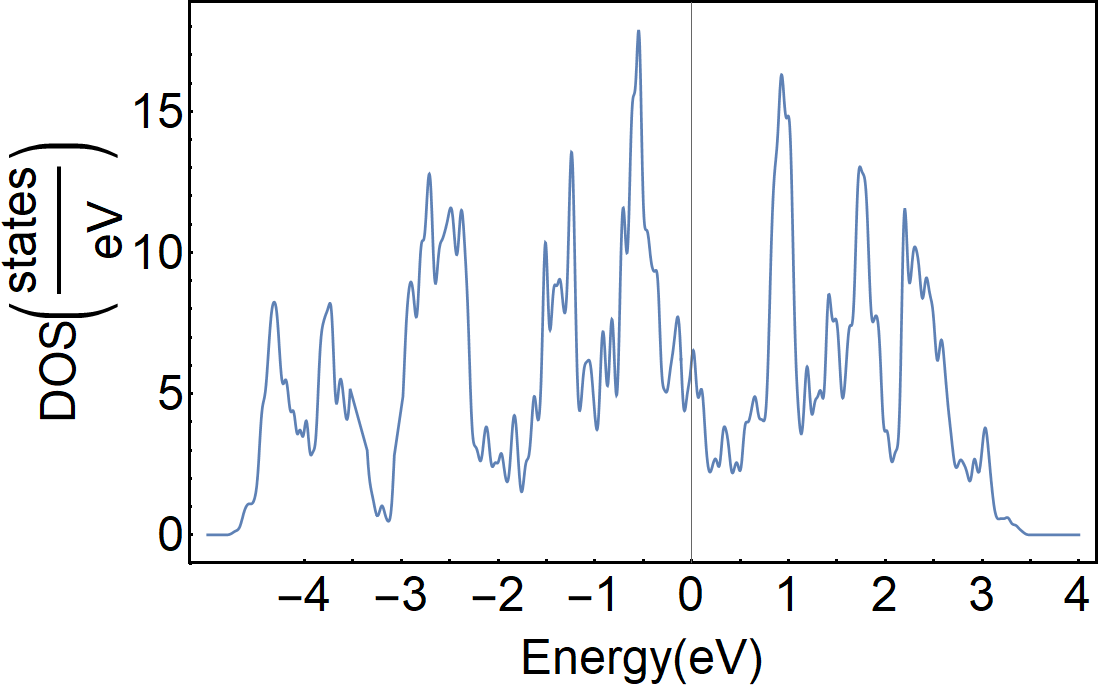} \caption{Total density of states of K$_{2}$Cr$_{3}$As$_{3}$ (Fermi level
is at zero energy).}
\label{fig:totdos} 
\end{figure}

In order to evaluate the orbital character of the low-energy excitations
around the Fermi level, we calculate the total density of states (DOS),
together with its projection on the Cr and As relevant orbitals. The
total DOS is obtained from the standard definition 
\begin{equation}
\rho(\epsilon)=\frac{1}{N}\sum_{\boldsymbol{k}}\delta(\epsilon-\epsilon_{\boldsymbol{k}})\label{eqn:DOS}
\end{equation}
in which $\epsilon_{\boldsymbol{k}}$ is the energy dispersion as
obtained from the tight-binding calculation presented in the previous
Section, and the sum is carried out on the Brillouin zone, our grid
consisting of $6\times6\times12$ $\boldsymbol{k}$ points. The delta
functions in Eq.~(\ref{eqn:DOS}) have been approximated by Gaussian
functions where the variance is assumed to be $\sigma=$ 0.025~eV.
The total DOS reported in Fig.~\ref{fig:totdos} exhibits, as expected,
peaks in correspondence of the flat portions of the energy spectrum.
Similarly to what found for CrAs~\citep{Autieri17,Autieri18,CAutieri17},
the DOS has a predominant As character at energies of the order of
$\pm2\,$eV away from the Fermi level. The peaks near -0.5 eV and
1 eV are instead associated with the Cr-$d_{yz}$ and $d_{xz}$ orbitals,
whereas, differently from what happens for CrAs, there is no clear
prevalence of Cr states around the Fermi energy, but rather the Cr-$d_{xy}$,
$d_{x^{2}-y^{2}}$ and $d_{z^{2}}$ and the As-$p_{x}$ and $p_{y}$
contributions are all relevant, as we will point out below in more
detail.

\begin{figure}
\centering \includegraphics[width=7cm]{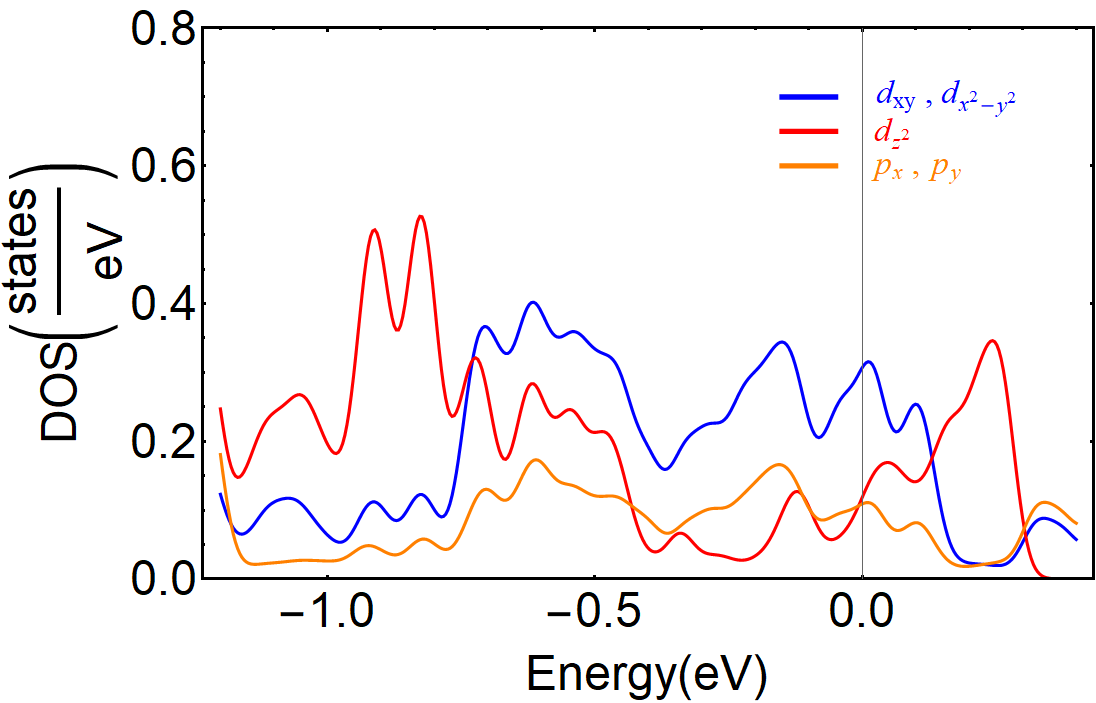} \caption{DOSs per atom of K$_{2}$Cr$_{3}$As$_{3}$ projected onto orbitals
symmetric with respect to the basal plane, i.e. $d_{xy}$ and $d_{x^{2}-y^{2}}$
(blue line), $d_{z^{2}}$ (red line) and $p_{x}$ and $p_{y}$ (orange
line). The curves have been obtained averaging the projected DOSs
over all the atoms of the unit cell.}
\label{fig:dsdensity} 
\end{figure}

\begin{figure}
\centering \includegraphics[width=7cm]{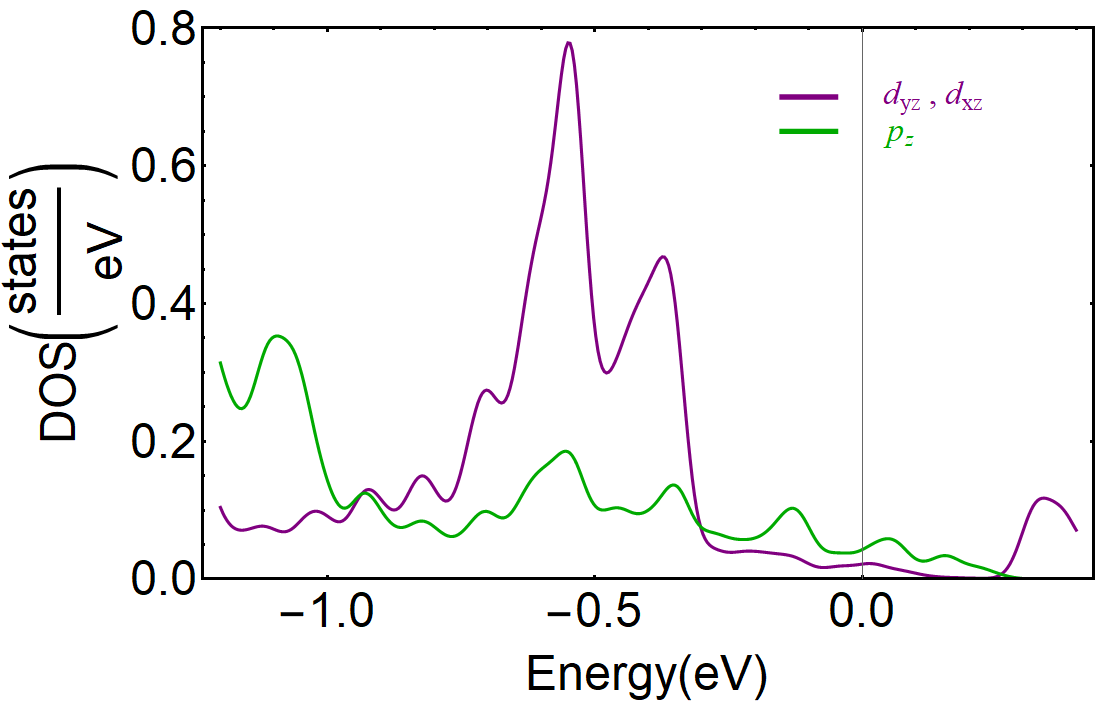} \caption{Same as in Fig.\ref{fig:dsdensity} for antisymmetric orbitals, i.e.
$d_{yz}$ and $d_{xz}$ (purple line) and $p_{z}$ (green line). }
\label{fig:pdsdensity} 
\end{figure}

We have also determined the projections of the total DOS on the orbitals
of the Cr or As atoms of the material. These are defined as 
\begin{equation}
\rho_{\alpha}(\epsilon)=\frac{1}{N}\sum_{\boldsymbol{k}}|\bigl\langle\psi_{\boldsymbol{k}}|f_{\alpha}\bigr\rangle|^{2}\delta(\epsilon-\epsilon_{\boldsymbol{k}})\label{eqn:DOSlocal}
\end{equation}
where $\psi_{\boldsymbol{k}}$ are the eigenstates of our problem
and $f_{\alpha}$ represents the orbital on which we project (the
delta functions are again approximated with Gaussians). The projected
DOSs associated with the orbitals symmetric with respect to the basal
plane, i.e. $d_{xy}$, $d_{x^{2}-y^{2}}$, $d_{z^{2}}$, $p_{x}$,
$p_{y}$, are shown in Fig.~\ref{fig:dsdensity}, while those associated
with the antisymmetric ones, i.e. $d_{yz}$, $d_{xz}$, $p_{z}$,
are presented in Fig.~\ref{fig:pdsdensity}. 

Their behavior confirms the results of first-principle calculations,
namely the orbitals that dominate the low-energy excitations are the
chromium $d_{xy}$, $d_{x^{2}-y^{2}}$ and $d_{z^{2}}$~\citep{Jiang15},
with the highest contribution corresponding to a pronounced peak at
the Fermi energy associated with the $d_{xy}$ and $d_{x^{2}-y^{2}}$
orbitals. Nonetheless, we see that an appreciable contribution also
comes from As $p_{x}$ and $p_{y}$ orbitals, this signaling the difficulty
of reducing the full Hamiltonian (\ref{eqn:tightbinding}) to a simpler
effective one where the $d$ and the $p$ orbital degrees of freedom
are efficiently disentangled. Finally, from Fig.~\ref{fig:pdsdensity}
we see that the projected DOS for antisymmetric orbitals exhibits
negligible contribution at the Fermi energy, providing evidence of
the decoupling between the two sectors corresponding to orbitals symmetric
or antisymmetric with respect to the basal plane.

\begin{figure}
\centering \includegraphics[width=7cm]{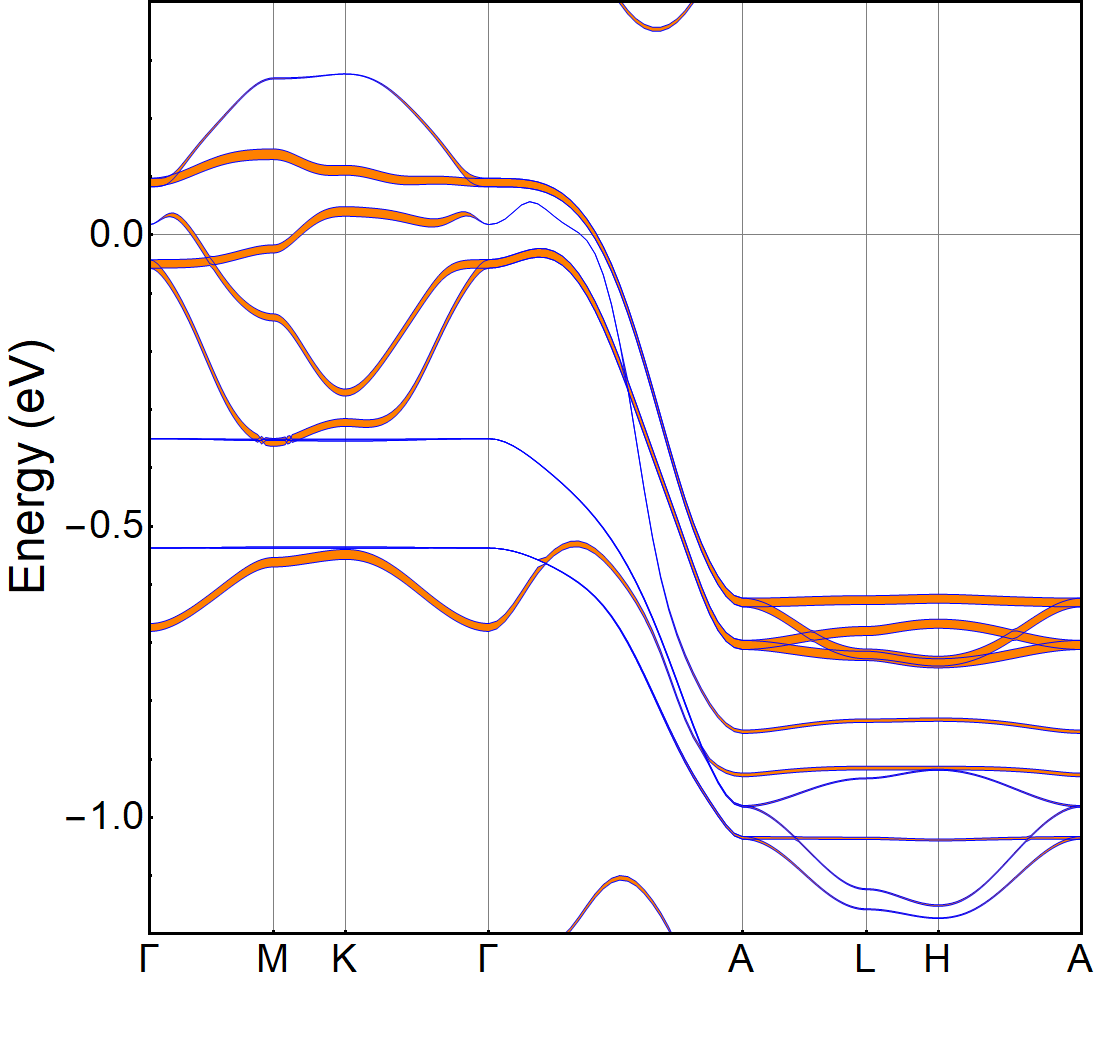} \caption{Contribution to the tight binding band structure (blue curve) of the
$d_{xy}$ and $d_{x^{2}-y^{2}}$ orbitals, represented as fat band
(in orange).}
\label{fig:b1} 
\end{figure}

\begin{figure}
\centering \includegraphics[width=7cm]{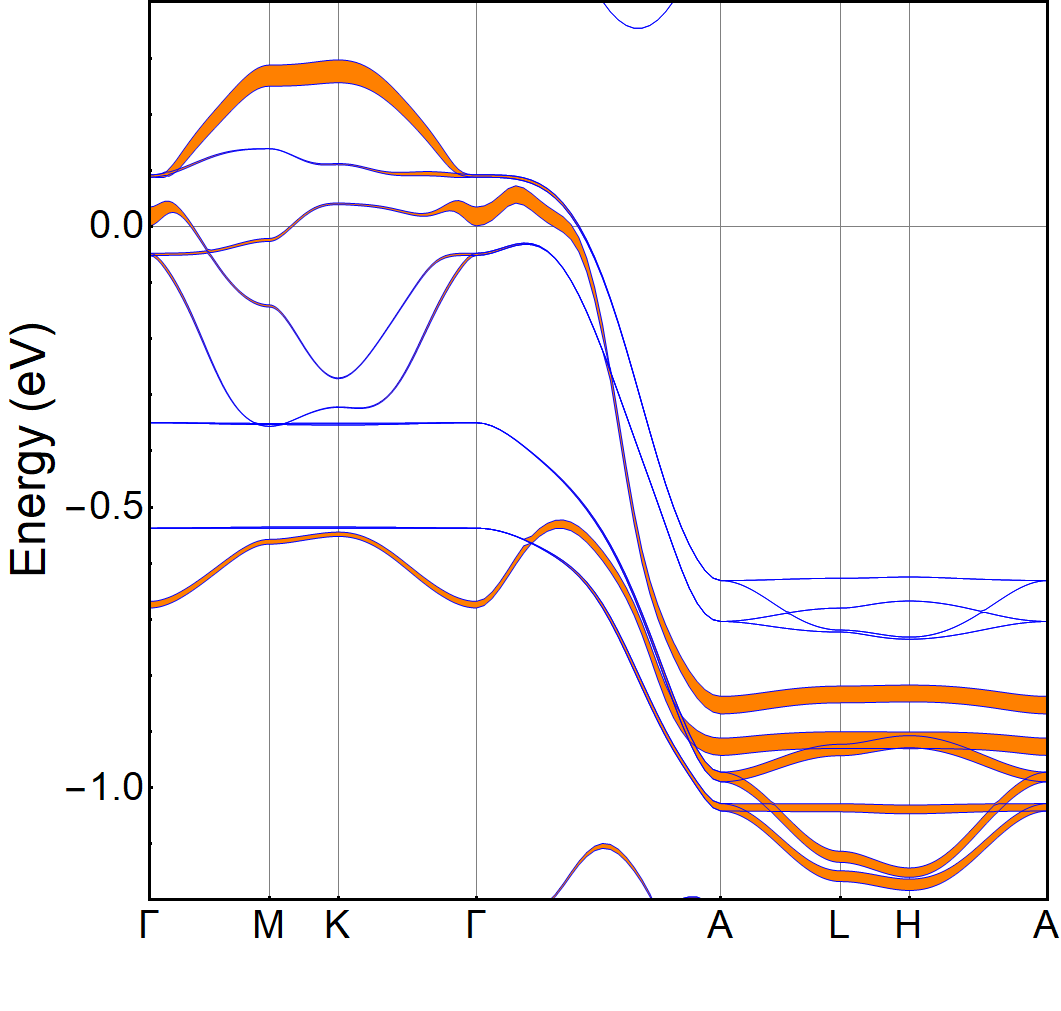} \caption{Same as in Fig.\ref{fig:b1}, referred to the $d_{z^{2}}$ orbital.}
\label{fig:b2} 
\end{figure}

\begin{figure}
\centering \includegraphics[width=7cm]{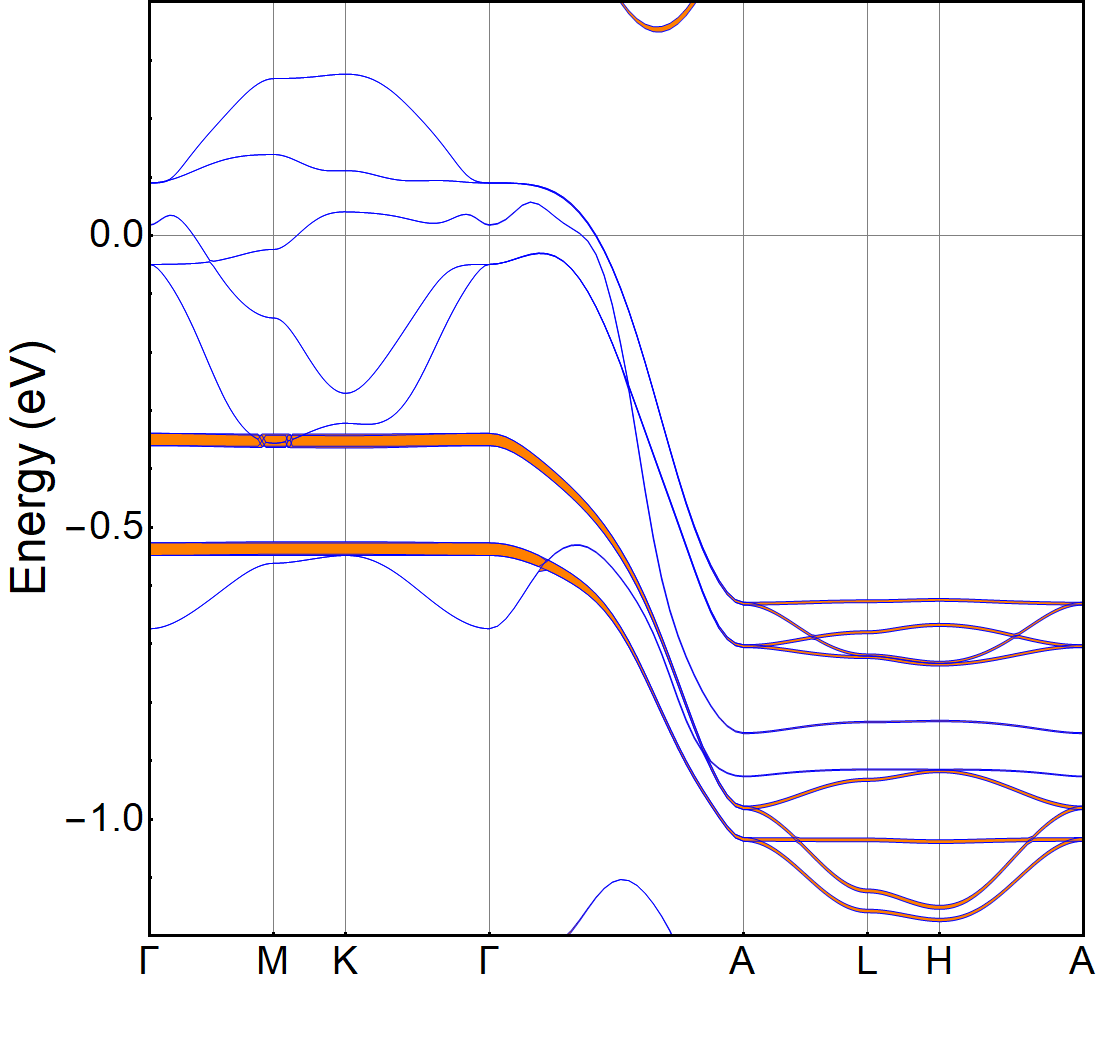} \caption{Same as in Fig.\ref{fig:b1}, referred to the $d_{xz}$ and $d_{yz}$
orbitals.}
\label{fig:b3} 
\end{figure}

\begin{figure}
\centering \includegraphics[width=7cm]{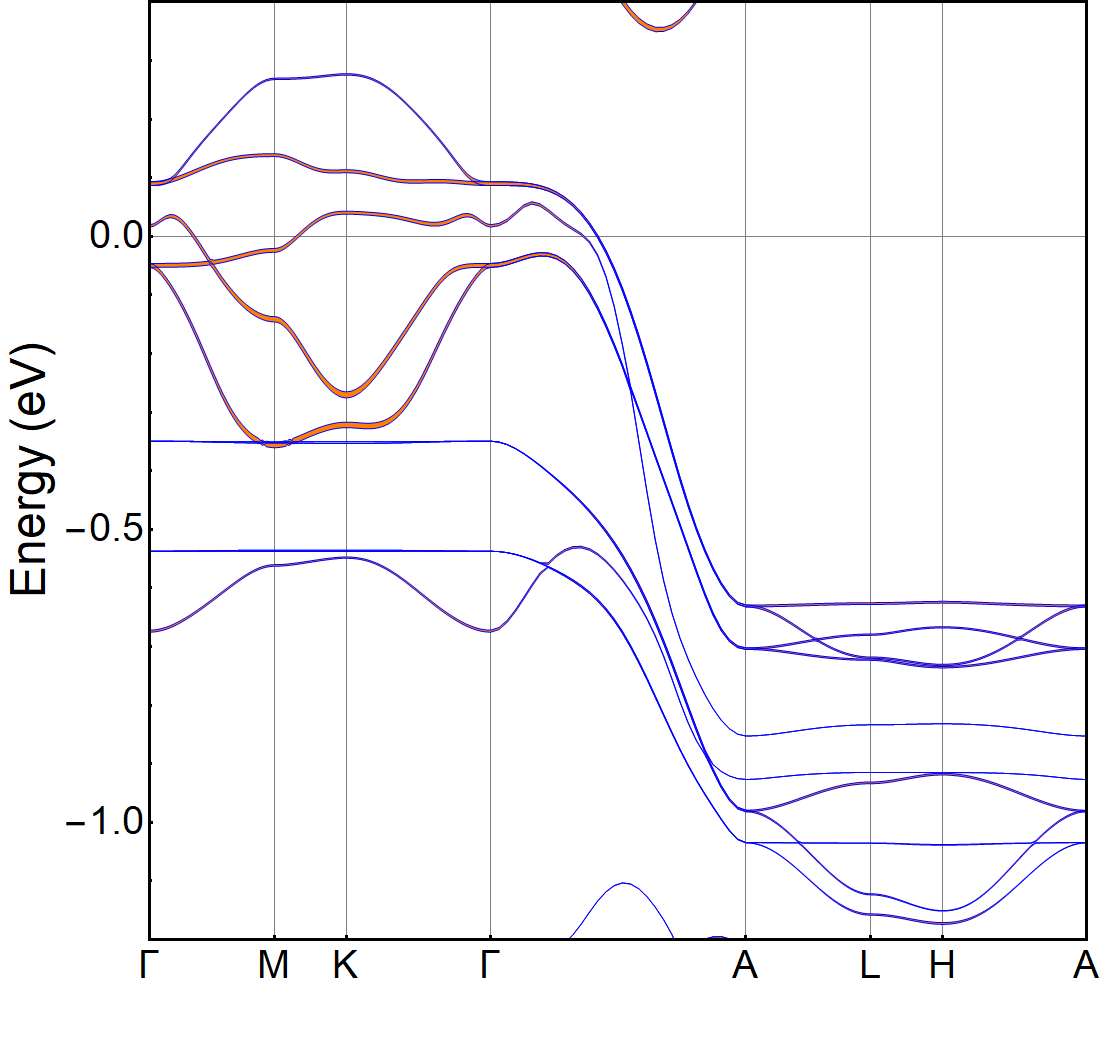} \caption{Same as in Fig.\ref{fig:b1}, referred to the $p_{x}$ and $p_{y}$
orbitals.}
\label{fig:b4} 
\end{figure}

\begin{figure}
\centering \includegraphics[width=7cm]{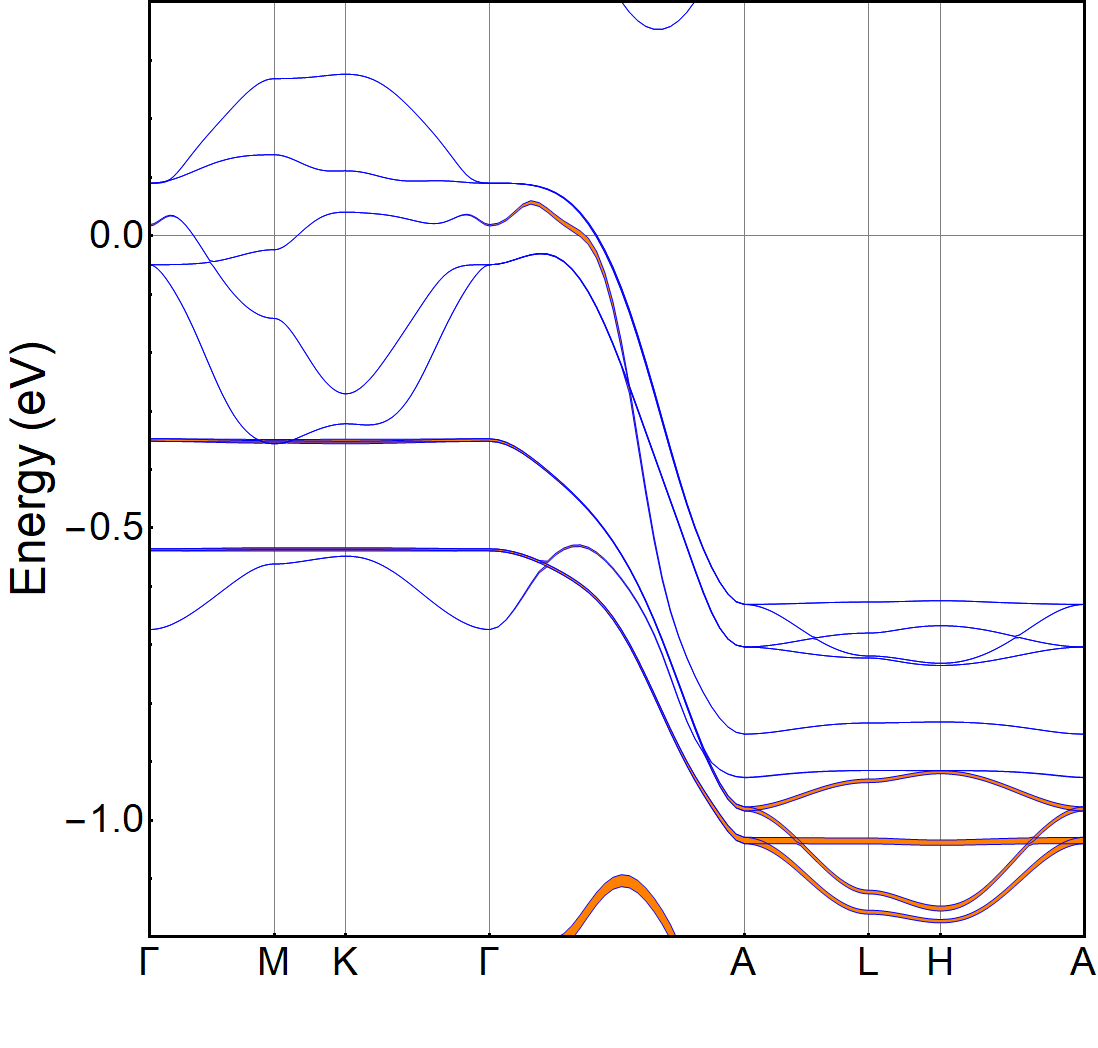} \caption{Same as in Fig.\ref{fig:b1}, referred to the $p_{z}$ orbital.}
\label{fig:b5} 
\end{figure}

To gain a better insight into the nature of the isolated set of ten
bands in the energy window {[}-1.2 eV, 0.4 eV{]} around the Fermi
level, we have performed a detailed analysis of the orbital character
of each energy level along the main directions in the Brillouin zone.
This is provided through the "fat bands" representation, where the
width of each band-line is proportional to the weight of the corresponding
orbital component, as shown in Figs.~\ref{fig:b1}-\ref{fig:b5}.
One can notice that an accurate description of the conduction and
valence bands along the various paths involves both Cr and As. As
one can see, the three bands crossing the Fermi level are mainly built
from the $d_{xy}$, $d_{x^{2}-y^{2}}$, $d_{z^{2}}$ orbitals of Cr,
with a degree of mixing which is highly dependent on the selected
path in the Brillouin zone.

\section{Löwdin procedure}

As our previous analysis suggests, the symmetric orbitals $d_{xy}$,
$d_{x^{2}-y^{2}}$, $d_{z^{2}}$, $p_{x}$ and $p_{y}$ dominate at
the Fermi level, so one can project out the low-lying degrees of freedom
using the Löwdin downfolding procedure~\citep{lowdin50}. This method
is based on the partition of a basis of unperturbed eigenstates into
two classes, related to each other by a perturbative formula giving
the influence of one class of states on the other one. In this case,
the two classes are the symmetric (s) and antisymmetric (a) $p$ and
$d$ orbitals with respect to the basal plane.

Schematically, given the basis defined in Eq.(\ref{eqn:vectors}),
the matrix has the structure 
\begin{equation}
H=\left[\begin{array}{c|c}
H_{ss} & H_{sa}\\
\hline H_{as} & H_{aa}
\end{array}\right]\,,\label{eqn:matrix}
\end{equation}
where $H_{ss}$ is the submatrix including hoppings between symmetric
orbitals, $H_{sa}$ hoppings between symmetric and anti-symmetric
orbitals, and $H_{aa}$ hoppings between anti-symmetric orbitals.

The submatrices are in turn made of block matrices. Considering for
instance $H_{ss}$, we have 
\begin{equation}
H_{ss}=\left[\begin{array}{c|c}
H_{Cr_{s}Cr_{s}} & H_{Cr_{s}As_{s}}\\
\hline H_{As_{s}Cr_{s}} & H_{As_{s}As_{s}}
\end{array}\right]\,,
\end{equation}
where the subscripts indicate the orbitals involved, so that 
\[
H_{Cr_{s}Cr_{s}}=\begin{bmatrix}H_{xy/xy} & H_{xy/x^{2}-y^{2}} & H_{xy/z^{2}}\\
H_{x^{2}-y^{2}/xy} & H_{x^{2}-y^{2}/x^{2}-y^{2}} & H_{x^{2}-y^{2}/z^{2}}\\
H_{z^{2}/xy} & H_{z^{2}/x^{2}-y^{2}} & H_{z^{2}/z^{2}}
\end{bmatrix},
\]
\[
H_{As_{s}As_{s}}=\begin{bmatrix}H_{x/x} & H_{x/y}\\
H_{y/x} & H_{y/y}
\end{bmatrix}.
\]
and similarly for $H_{Cr_{s}As_{s}}$. Here $xy$, $x^{2}-y^{2}$,
$z^{2}$, $yz$ and $xz$ denote the five $d$ orbitals of the Cr
atoms, while $x$, $y$ and $z$ denote the three $p$ orbitals of
the As atoms. For example, $H_{xy/x^{2}-y^{2}}$ is the submatrix
that includes all the hopping processes between the $d_{xy}$ and
$d_{x^{2}-y^{2}}$ orbitals belonging to the six chromium atoms.

\begin{figure}
\centering \includegraphics[width=7cm]{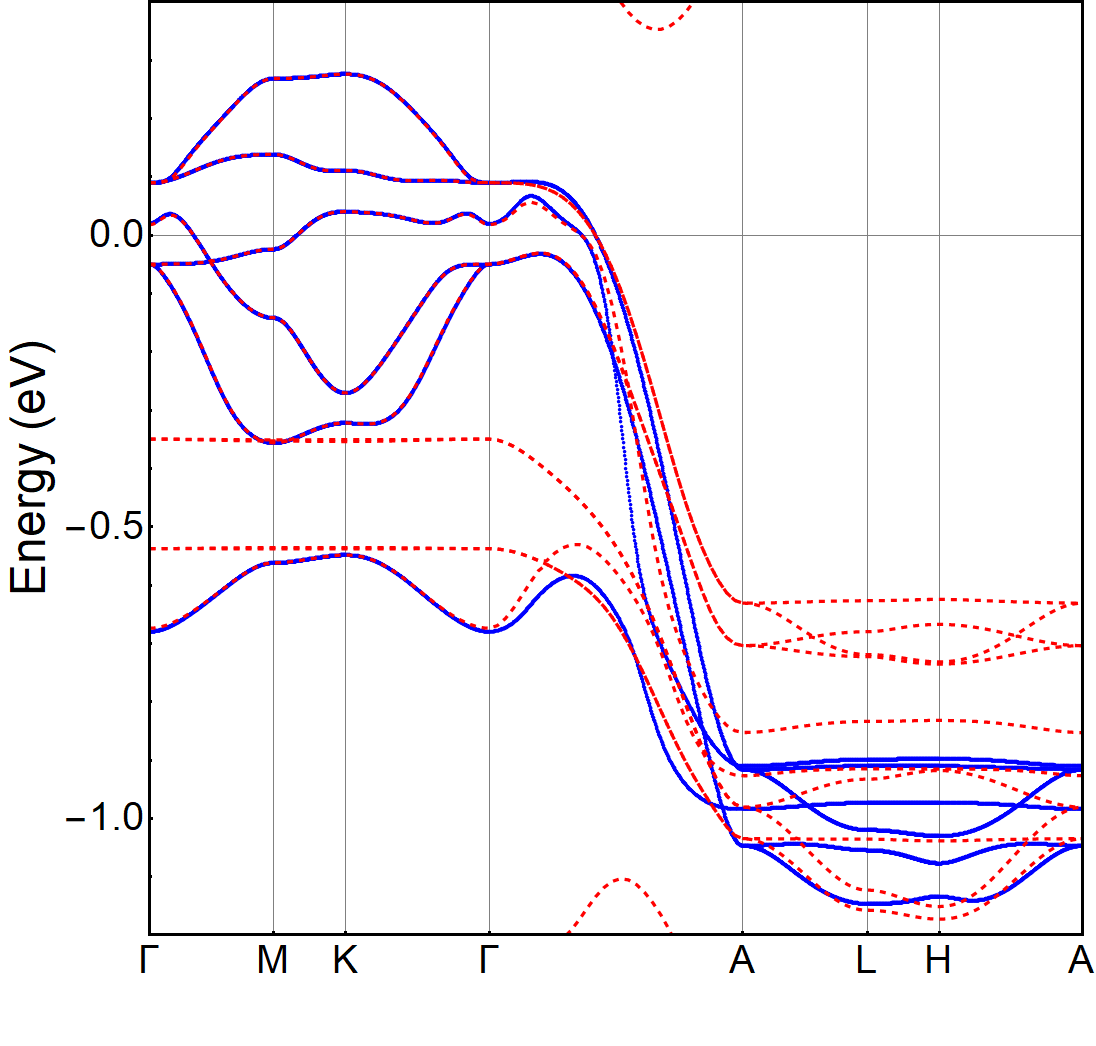} \caption{Comparison in the energy range around the Fermi level between the
DFT band structure (red dashed lines) and the one obtained from the
Löwdin downfolding procedure described in the text (blue lines).}
\label{fig:Lowdin1} 
\end{figure}

Referring to the matrix of Eq.~(\ref{eqn:matrix}) and downfolding
the $H_{aa}$ submatrix, the solution of the original eigenvalue problem
is mapped to that of a corresponding effective Hamiltonian $\widetilde{H}_{ss}$,
whose rank is 30, with $\widetilde{H}_{ss}$ given by~\citep{andersen95}
\begin{equation}
\widetilde{H}_{ss}(\varepsilon)={H}_{ss}-{H}_{sa}\left({H}_{aa}-\varepsilon\mathbb{I}\right)^{-1}{H}_{as}\,.\label{eqn:equationL}
\end{equation}
Using this technique, we get the low-energy effective Hamiltonian
projected into the subsector given by the symmetric orbitals, going
beyond the simpler complete Wannier function method. The band structure
that we have obtained applying the Löwdin procedure is shown in Fig.~\ref{fig:Lowdin1},
where the DFT spectrum near the Fermi level is also reported for comparison.
We can see that the band structure near the Fermi level is caught
to a high degree of approximation and the agreement is almost complete.
The $d$ and $p$ anti-symmetric orbitals thus can be fully disentangled
from the symmetric ones, as a consequence of the peculiar geometry
corresponding to the arrangement of the chromium atoms.

It is worth noting that we still have a disagreement in the A-L-H-A
region. Such an occurrence is due to the predominance of the weight
of the $d_{z^{2}}$ orbital in that region of the Brillouin zone,
although the corresponding bands are somehow distant from the Fermi
surface.

\section{Derivation of a minimal five-band tight-binding model}

On the basis of the indications provided by the orbital characterization
of the band structure and by the Löwdin procedure, we now introduce
a minimal tight-binding model allowing to satisfactorily reproduce
the energy spectrum around the Fermi energy in the whole ${\bf k}$-space.
We start by referring to the isolated set of ten bands developing
in the energy range going approximately from -1.2 to 0.4 eV (see,
for instance, Fig.~\ref{fig:Lowdin1}). The fat band representation
used in Figs.~\ref{fig:b1}-\ref{fig:b5} provides evidence that
these bands have mainly the character of the orbitals that are symmetric
with respect to the basal plane. The Löwdin projection clearly demonstrated
that downfolding the ten bands over the six symmetric ones, it is
possible to obtain a very good description of the energy bands in
proximity of the Fermi energy. These results naturally suggest a further
refinement of our calculations, consisting in an application of the
Wannier method taking explicitly into account the predominant weight
of the symmetric states. We eventually find that this combination
of the Löwdin and the Wannier approaches allows to obtain a fully
reliable minimal tight-binding model.

We observe that the non-dispersive bands in the $k_{z}$=0 plane present
an anti-symmetric character with weight mainly coming from $d_{xz}$
and $d_{yz}$ orbitals. Moreover, as one can see from the behavior
of the DOS shown in Fig.~\ref{fig:pdsdensity}, the contribution
at the Fermi level of these bands, as well as the one of the $p$
bands, is small compared to that of the symmetric ones. This suggests
to exclude the anti-symmetric bands from the construction of a simplified
model Hamiltonian, and thus to consider only the six symmetric ones,
associated with two $d_{xy}$, two $d_{x^{2}-y^{2}}$ and two $d_{z^{2}}$
orbitals. A further simplification is applied limiting to one the
number of the $d_{z^{2}}$ orbital, in consideration of the fact that
the corresponding band is the one lying farther from the Fermi energy.
We thus perform the Wannier calculation referring to a five-band effective
model, consistently with the fact that four bands cut the Fermi level,
one of them being doubly degenerate at $\Gamma$ point.

\begin{figure}
\centering \includegraphics[width=7cm]{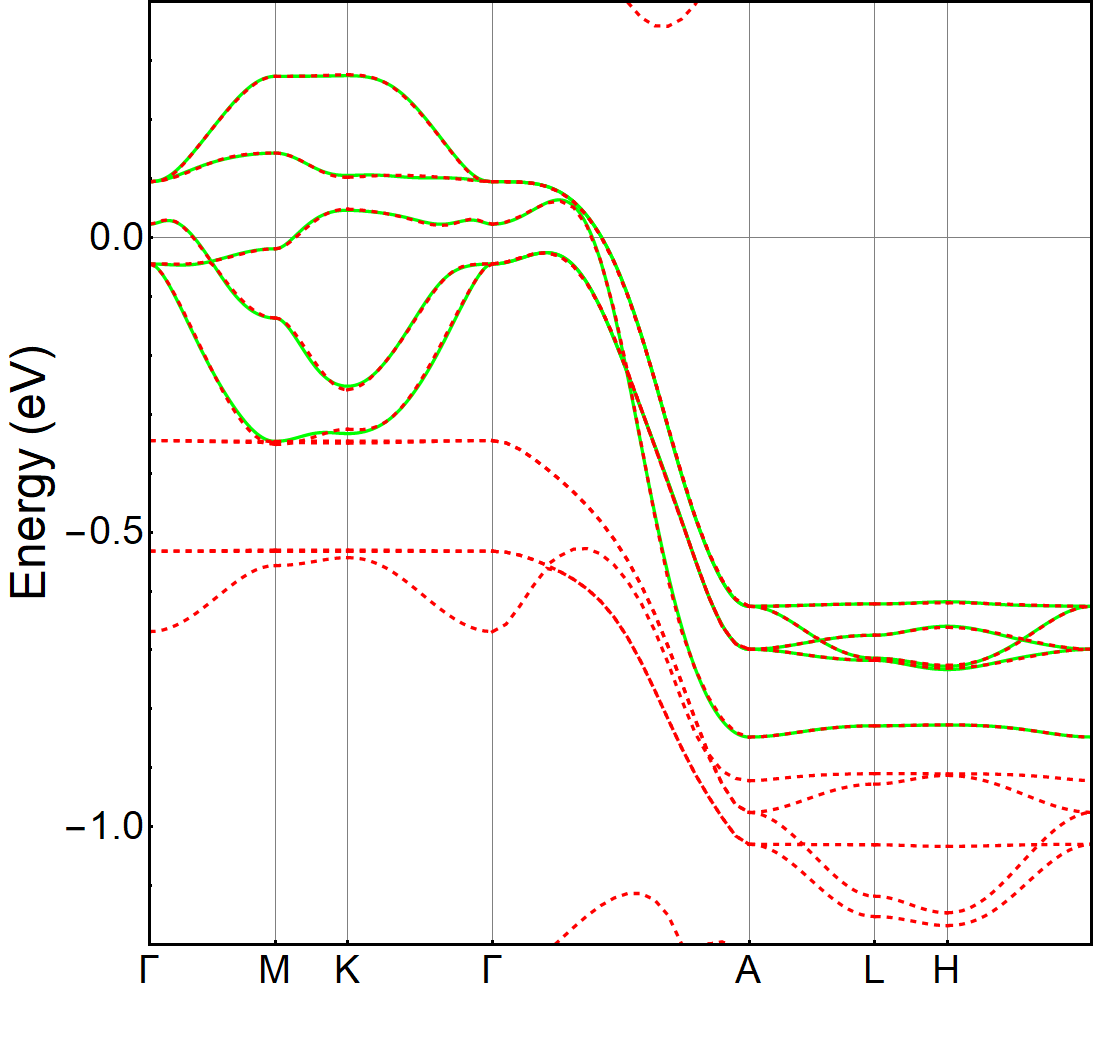} \caption{Comparison between the LDA band structure (red dashed lines) and the
one obtained from the five-band model described in the text (green
solid lines).}
\label{fig:minimal} 
\end{figure}

Since chromium-based compounds, such as K$_{2}$Cr$_{3}$As$_{3}$,
exhibit weak or moderate electronic correlations, they have a covalent
character rather than a ionic one, so that, in the case of low-dimensional
systems, a Wannier function can also be placed between equivalent
atoms~\citep{Autieri17a}. Our choice is to place $d_{xy}$, $d_{x^{2}-y^{2}}$,
and $d_{z^{2}}$ wave functions in the middle of the Cr-triangle belonging
to the KCr$_{3}$As$_{3}$ plane, locating the other two $d_{xy}$
and $d_{x^{2}-y^{2}}$ wave functions in the middle of the Cr-triangle
lying in the K$_{3}$Cr$_{3}$As$_{3}$ plane. The interpolated band
structure obtained by this method is shown in Fig.~\ref{fig:minimal}
together with the DFT band structure. We can observe a perfect match
between the two spectra, thus demonstrating that our five-band model
allows to describe the low energy physics in a range of about 0.3
eV around the Fermi level with the same accuracy provided by DFT.
We also notice that the mixing of two different types of orbital which
is at the basis of the model, suggests that K$_{2}$Cr$_{3}$As$_{3}$
might actually behave as a two-channel Stoner $d$-electron metallic
magnet~\citep{Wysokinski2018}. Interestingly, this effect can drive
pressure-induced transitions between ferromagnetic and antiferromagnetic
ground states. 

We now derive the analytic expression of our tight-binding model,
including in the calculation three NN hopping terms along $z$, one
in the plane and one along the diagonal. We will denote by $\alpha_{1}$
and $\alpha_{2}$ the Wannier functions relative to the orbitals in
the plane at $z=c/2$ with predominant $xy$ and $x^{2}-y^{2}$ character,
$c/2$ being the distance between KCr$_{3}$As$_{3}$ and K$_{3}$Cr$_{3}$As$_{3}$
planes, and by $\alpha_{3}$, $\alpha_{4}$ and $\alpha_{5}$ those
relative to the orbitals with predominant $xy$, $x^{2}-y^{2}$ and
$z^{2}$ character, respectively, in the plane with $z=0$. We will
also denote by $t_{\alpha_{i},\alpha_{j}}^{lmn}$ the hopping amplitudes
between the Wannier states $\alpha_{i}$ and $\alpha_{j}$ along the
direction l$\mathbf{x}$ + m$\mathbf{y}$ + n$\mathbf{z}$. Since
the system exhibits inversion symmetry along the $z$ axis and the
orbitals under consideration are even, we will get terms proportional
to $cos(nk_{z}c/2)$ for the hopping along $z$, $n$ being an even
(odd) integer for hopping between homologous (different) orbitals.

According to the above assumptions, the Hamiltonian in momentum space
is represented as a 5$\times$5 matrix, with elements $H_{\alpha_{i},\alpha_{j}}$.
Concerning the diagonal elements, i.e. those referring to the same
Wannier state, they result from the sum of different contributions
related to on-site, out-of-plane and in-plane amplitudes, respectively.
They thus read as 
\[
H_{\alpha_{i},\alpha_{i}}(k_{x},k_{y},k_{z})=H_{\alpha_{i},\alpha_{i}}^{0}+H_{\alpha_{i},\alpha_{i}}^{\perp}(k_{z})+H_{\alpha_{i},\alpha_{i}}^{\parallel}(k_{x},k_{y})
\]
where 
\begin{eqnarray*}
H_{\alpha_{i},\alpha_{i}}^{0} & = & t_{\alpha_{i},\alpha_{i}}^{000}\equiv\varepsilon_{\alpha_{i}}^{0}\\
H_{\alpha_{i},\alpha_{i}}^{\perp}(k_{z}) & = & \sum_{n=1,2,3}2t_{\alpha_{i},\alpha_{i}}^{00n}\cos{(nk_{z}c)}\\
H_{\alpha_{i},\alpha_{i}}^{\parallel}(k_{x},k_{y}) & = & 2t_{\alpha_{i},\alpha_{i}}^{100}\cos{(k_{x}a)}\\
 &  & +4t_{\alpha_{i},\alpha_{i}}^{010}\cos{(k_{x}\frac{a}{2})}\cos{(k_{y}a\frac{\sqrt{3}}{2})}\;,
\end{eqnarray*}
with the numerical values of the hopping parameters being reported
in Table~\ref{TabDIAG}.

\noindent 
\begin{table}[t!]
\begin{centering}
\begin{tabular}{|c|c|c|c|c|c|c|}
\hline 
%{|p{16.0cm}|}
 & \multicolumn{1}{c|}{on site} & \multicolumn{3}{c|}{out of plane} & \multicolumn{2}{c|}{in plane}\tabularnewline
\hline 
 & 000  & 001  & 002  & 003  & 100  & 010 \tabularnewline
\hline 
$\alpha_{1}$  & -86.6  & 154.1  & -53.0  & -6.3  & 23.0  & -3.5 \tabularnewline
\hline 
$\alpha_{2}$  & -86.6  & 154.1  & -53.0  & -6.3  & -12.3  & 14.2 \tabularnewline
\hline 
$\alpha_{3}$  & -37.0  & 165.0  & -41.9  & -2.9  & 30.9  & -0.2 \tabularnewline
\hline 
$\alpha_{4}$  & -37.0  & 165.0  & -41.9  & -2.9  & -10.6  & 20.6 \tabularnewline
\hline 
$\alpha_{5}$  & 0  & 271.6  & -63.9  & -14.2  & -15.4  & -15.4 \tabularnewline
\hline 
\end{tabular}
\par\end{centering}
\caption{On-site energies and out-of-plane and in-plane hopping integrals between
the same Wannier states. The on-site energy of the $z^{2}$-like function
is set to zero (energy units in meV).}
\label{TabDIAG} 
\end{table}

Going to the off-diagonal elements connecting different Wannier states,
we first observe on a general ground that when a crystal structure
exhibits a reflection symmetry with respect to the $x$ axis, one
has for pure $d$-orbitals $t_{xy,x^{2}-y^{2}}^{100}=0$ and $t_{xy,x^{2}-y^{2}}^{010}=t_{xy,x^{2}-y^{2}}^{0\bar{1}0}$.
As regards K$_{2}$Cr$_{3}$As$_{3}$, we have that its crystal structure
is symmetric with respect to the $y$-axis, but not with respect to
the $x$-axis. Since the Wannier functions keep this missing symmetry,
we have $t_{\alpha_{1},\alpha_{2}}^{100}\neq0$ and $t_{\alpha_{1},\alpha_{2}}^{010}\neq t_{\alpha_{1},\alpha_{2}}^{0\bar{1}0}$.
We stress that in our tight-binding model this effect is explicitly
taken into account, differently from previous approaches where the
above-mentioned $x$-axis symmetry is nonetheless applied~\citep{XWu15,Zhang16}.
As in the previous case, we have that the non-diagonal elements of
the Hamiltonian result from in-plane and out-of-plane contributions
associated with hopping processes connecting different Wannier states.
Their expressions are reported in Appendix B, together with the Tables
giving the numerical values of the hopping amplitudes involved.

\section{Conclusions}

We have presented a method that combines the Löwdin and the Wannier
procedures to derive a minimal five-band tight-binding model correctly
describing the low-energy physics of K$_{2}$Cr$_{3}$As$_{3}$ in
terms of four planar orbitals ($d_{xy}$ and $d_{x^{2}-y^{2}}$ for
each of the two planes KCr$_{3}$As$_{3}$ and K$_{3}$Cr$_{3}$As$_{3}$)
and a single out-of-plane one ($d_{z^{2}}$). We are confident that
this combined method can be applied to other transition-metal compounds,
including the iron-based superconductors.

Our results give clear indication that the physics of the system is
significantly affected by in-plane dynamics, in spite of the presence
in the lattice of well-defined quasi-1D nanotube structures. The results
presented here also make evident the minor role played by the local
electronic correlations in determining the physical properties of
the compound. Indeed, the inclusion within a LDA+U calculation scheme
of a non-vanishing Hubbard repulsion developing in the Cr $d$-orbitals
leads to only slight quantitative differences with respect to the
non-interacting case.

We notice that although a six-band model was previously reported~\citep{Zhang16}
using six symmetric orbitals, the five-band model proposed here describes
with higher accuracy the low-energy physics as a consequence of the
application of the Wannier method. We also point out that, with a
filling of four electrons shared among two kinds of orbitals, the
planar $d_{xy}$ and $d_{x^{2}-y^{2}}$ and the out-of-plane $d_{z^{2}}$
ones, the system might be in the Hund's metal regime. In this framework,
it has been proposed that Hund's coupling may lead to an orbital decoupling
that makes the orbitals independent from each other, so that some
of them can acquire a remarkably larger mass enhancement with respect
to the other ones. Furthermore, a possible connection between the
orbital-selective correlations and superconductivity might be investigated:
the selective correlations could be the source of the pairing glue
or, alternatively, could strengthen the superconducting instability
arising from a more conventional mechanism based on the exchange of
bosons or spin fluctuations. Work in this direction is in progress.

Finally, we point out that the model may be used to study transport
properties, magnetic instabilities, as well as superconductivity in
anisotropic crystal structures~\citep{Aperis}, also allowing to
investigate dynamical effects in this class of superconductors~\citep{Demedici}.
In this case, the evidence that the main features of the energy spectrum
around the Fermi level are essentially determined by the three symmetric
$d_{xy}$, $d_{x^{2}-y^{2}}$ and $d_{z^{2}}$ Cr orbitals and by
the $p_{x}$ and $p_{y}$ As ones, provides a constraint on the form
of the superconducting order parameter that should be assumed in the
development of the theory.

\section*{Acknowledgments}

The work is supported by the Foundation for Polish Science through
the IRA Programme co-financed by EU within SG OP. C.A. was supported
by CNR-SPIN via the Seed Project CAMEO. C.A. acknowledges the CINECA
award under the ISCRA initiative IsC54 \textquotedblleft CAMEO\textquotedblright{}
Grant, for the availability of high-performance computing resources
and support.

\appendix

\section{Tight-binding parametrization}

In this appendix, we report the systematic procedure leading to the
TB parametrization based on the 48 atomic orbitals taken into account
in the model Hamiltonian (2). In the following, we perform the calculations
by first considering the hopping processes within the quasi one-dimensional
{[}(Cr$_{3}$As$_{3}$)$^{2-}${]}$^{\infty}$ double-walled nanotubes
only, and then including step by step inter-tube and longer-range
intra-tube processes.

\begin{figure}
\centering \includegraphics[width=8cm]{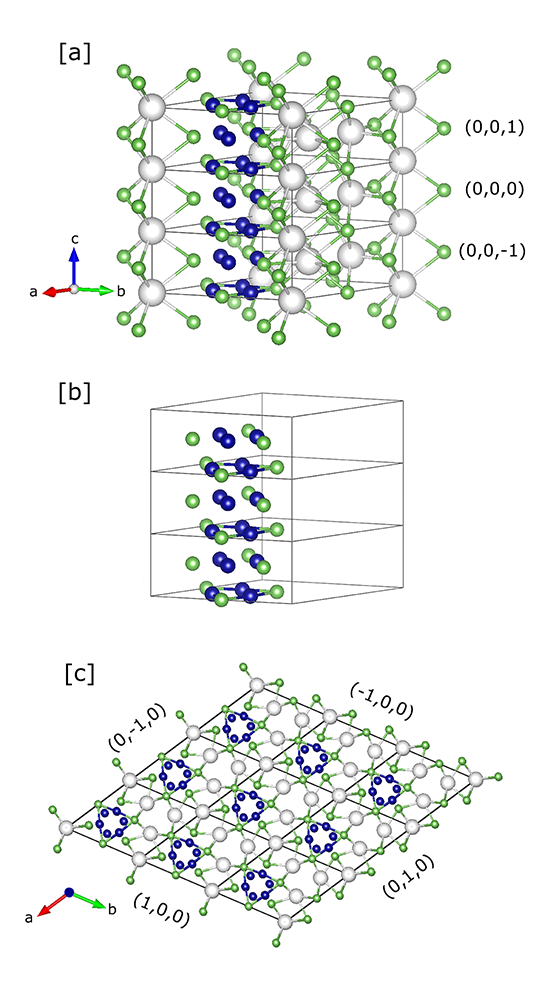} %	\caption{(a) (0,0,0), (0,0,1) and (0,0,-1) primitive cells of K$_2$Cr$_3$As$_3$. (b) The atoms of the sub-nanotube we consider at this level of approximation. (c) (0,0,0), (1,0,0), (-1,0,0), (0,1,0), (0,-1,0), (1,-1,0), (-1,1,0), (1,1,0) and (-1,-1,0) primitive cells.}
 \caption{(a) (0,0,0), (0,0,1) and (0,0,-1) primitive cells of K$_{2}$Cr$_{3}$As$_{3}$.
(b) Atoms taken into account in the diagonalization procedure when
only short-range intratube hopping processes are considered. (c) (0,0,0),
(1,0,0), (-1,0,0), (0,1,0), (0,-1,0), (1,-1,0), (-1,1,0), (1,1,0)
and (-1,-1,0) primitive cells.}
\label{fig:K2}
\end{figure}

\begin{figure}
\centering \includegraphics[width=7cm]{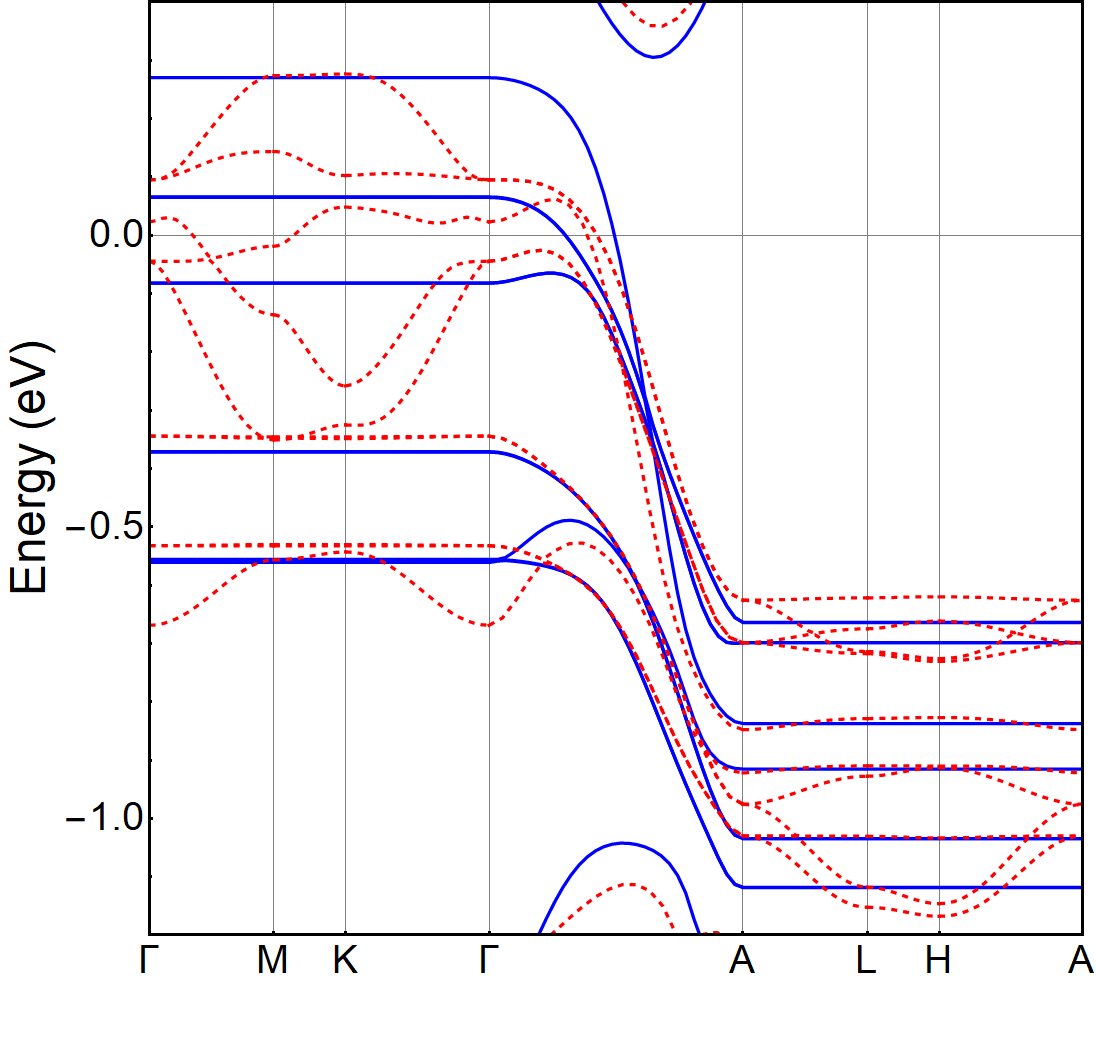} \caption{Comparison between the tight-binding band structure obtained considering
hoppings within the (0,0,0), (0,0,1) and (0,0,-1) primitive cells
(blue lines) and the DFT band structure (red dashed lines), in an
energy range around the Fermi level (set equal to zero).}
\label{fig:shortrangezoom}
\end{figure}

\subsection{Short range intra-tube hybridizations}

As previously pointed out, the most relevant sub-geometry of the K$_{2}$Cr$_{3}$As$_{3}$
lattice is a quasi-one dimensional double-walled sub-nanotube extending
mainly along the $z$-axis. So, if we consider only hopping processes
between intra-tube atoms, an already reasonable approximation of the
band structure can be obtained, in particular along the line of the
Brillouin zone associated with variations of $k_{z}$, i.e. the $\Gamma$-A
line. Referring to the notation $\bm{R}=n_{1}\bm{a}_{1}+n_{2}\bm{a}_{2}+n_{3}\bm{a}_{3}$,
we start by limiting ourselves to the primitive cells denoted by ($n_{1}$,$n_{2}$,$n_{3}$)=(0,0,0),
(0,0,1) and (0,0,-1), as shown in Fig.~\ref{fig:K2}(a-b).

The band structure that we have obtained is shown in Fig.~\ref{fig:shortrangezoom}
in an energy window around the Fermi level, with the DFT spectrum
being also reported for comparison. We can see that the bands are
flat along the in-plane paths of the Brillouin zone where $k_{x}$
and $k_{y}$ vary, as expected, but along the $\Gamma$-A line they
exhibit a behavior quite close to the DFT results. It is also evident
that more reliable results require in any case the inclusion of hopping
processes involving longer range intra-tube cells as well as inter-tube
ones.

\begin{figure}
\centering \includegraphics[width=7cm]{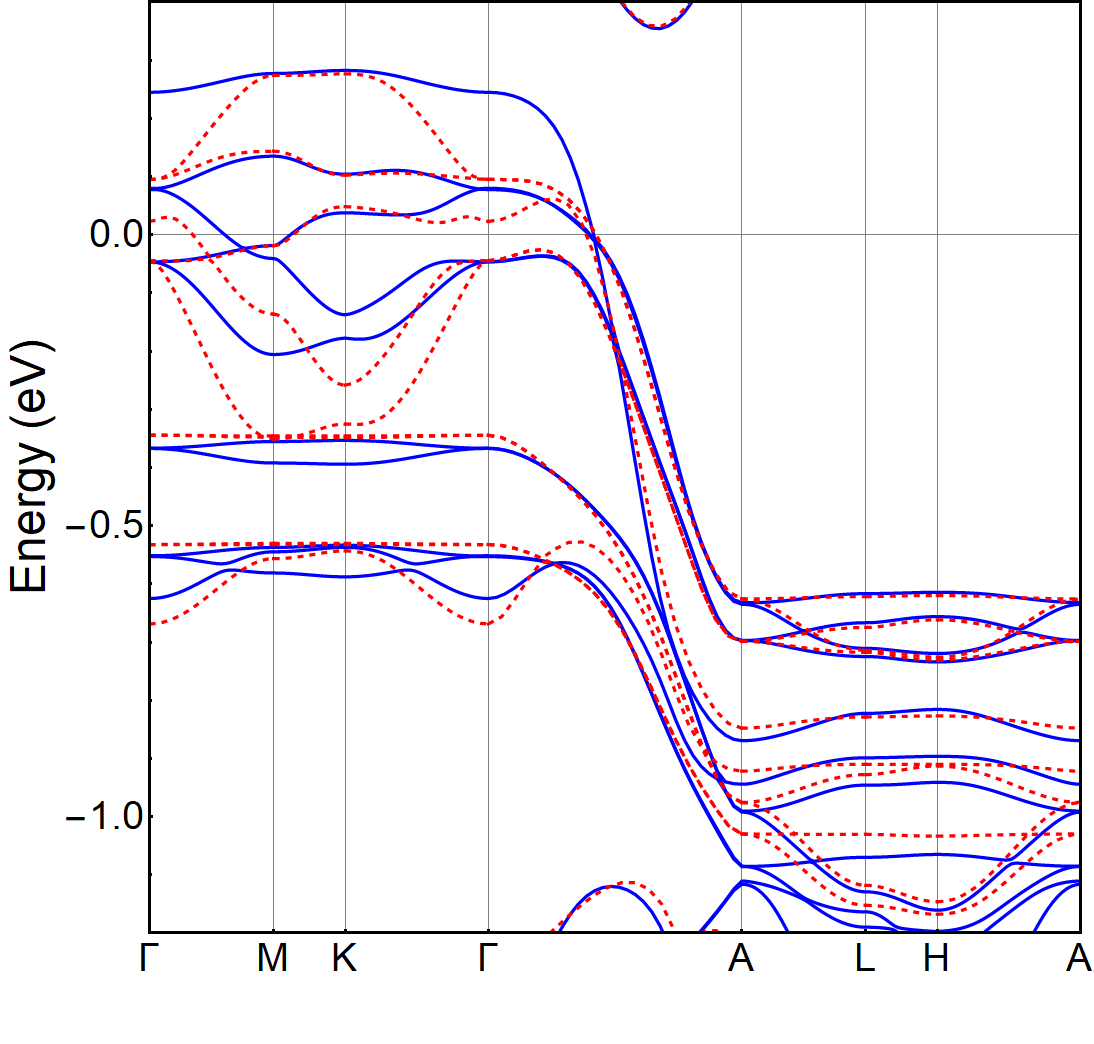}
%\caption{Comparison between the tight-binding band structure (blue lines) obtained considering the (0,0,0), %(0,0,1), (0,0,-1), (1,0,0), (-1,0,0), (0,1,0), (0,-1,0), (1,-1,0), (-1,1,0), (1,1,0) and (-1,-1,0) primitive cells %and the DFT band structure (red dashed lines) near the Fermi level.}
 \caption{Same as in Fig.~\ref{fig:shortrangezoom}, with the tight-binding
calculations extended to hopping processes in the (1,0,0), (-1,0,0),
(0,1,0), (0,-1,0), (1,-1,0), (-1,1,0), (1,1,0) and (-1,-1,0) primitive
cells.}
\label{fig:shortrangezoom_b}
\end{figure}

\begin{figure}
\centering \includegraphics[width=7cm]{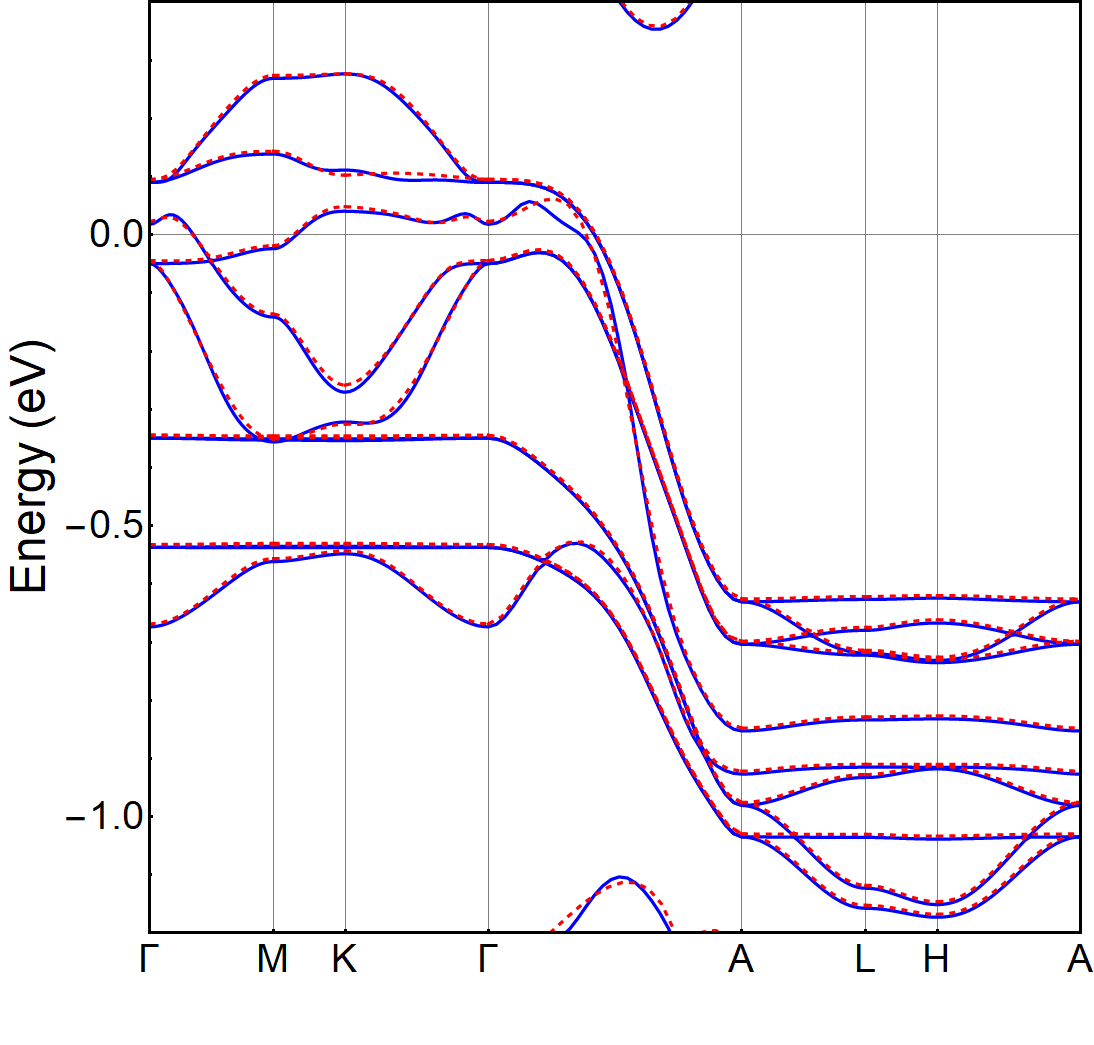} %\caption{Comparison between the tight-binding band structure (blue lines) obtained considering the hoppings until the second neighbour cells along the $z$-axis ($n_{3}=2$ and $n_{3}=-2$) and the in-plane interactions at the cells until $n_{3}=1$ and $n_{3}=-1$ and the DFT band structure (red dashed lines) near the Fermi level, that is set at zero energy.}
\caption{Same as in Fig.~\ref{fig:shortrangezoom_b}, with the tight-binding
calculations further extended to fifth neighbour cells along the$z$-axis
(from $n_{3}=5$ to $n_{3}=-5$) and to second-neighbor in-plane cells.}
\label{fig:longrangezoomz}
\end{figure}

\subsection{Inter-tube and long-range intra-tube hybridizations}

We now include in the diagonalization procedure inter-tube hoppings
in the $x$-$y$ plane ($n_{3}=0$), taking into account the contributions
coming from the cells (1,0,0), (-1,0,0), (0,1,0), (0,-1,0), (1,-1,0),
(-1,1,0), (1,1,0) and (-1,-1,0) (see Fig.~\ref{fig:K2}(c)). The
comparison of the band structure correspondingly obtained with the
one given by DFT (see Fig.~\ref{fig:shortrangezoom_b}) makes evident
that the agreement improves along the $\Gamma$-A line as well as
along the other lines of the Brillouin zone, though there are still
some qualitative differences, also at the Fermi level. In order to
get a truly satisfactory agreement, it is necessary to include all
hopping processes up to the fifth-neighbor cells along the z-axis
(from $n_{3}=5$ to $n_{3}=-5$), together with the in-plane hoppings
up to the second-neighbor cells (see Fig.~\ref{fig:longrangezoomz}).

\section{Off-diagonal elements of tight-binding Hamiltonian}

We report here the expressions of the off-diagonal elements $H_{\alpha_{i},\alpha_{j}}(k_{x},k_{y},k_{z})$
$(\alpha_{i}\ne\alpha_{j})$ of the tight-binding Hamiltonian introduced
in Section VI. They refer to hopping processes which connects different
Wannier states and have the following form:

\begin{eqnarray*}
H_{\alpha_{1},\alpha_{3}} & = & \sum_{n=0,1,2}2t_{\alpha_{1},\alpha_{3}}^{00n+\frac{1}{2}}\cos{((n+\frac{1}{2})k_{z}c)}\\
 &  & +4t_{\alpha_{1},\alpha_{3}}^{100}\cos{(k_{x}a)}\cos{(k_{z}c/2)}\\
 &  & +4t_{\alpha_{1},\alpha_{3}}^{010}e^{i(k_{y}a\frac{\sqrt{3}}{2})}\cos{(k_{x}a/2)}\cos{(k_{z}c/2)}\\
 &  & +4t_{\alpha_{1},\alpha_{3}}^{0\bar{1}0}e^{-i(k_{y}a\frac{\sqrt{3}}{2})}\cos{(k_{x}a/2)}\cos{(k_{z}c/2)}\\
H_{\alpha_{2},\alpha_{4}} & = & \sum_{n=0,1,2}2t_{\alpha_{2},\alpha_{4}}^{00n+\frac{1}{2}}\cos{((n+\frac{1}{2})k_{z}c)}\\
 &  & +4t_{\alpha_{2},\alpha_{4}}^{100}\cos{(k_{x}a)}\cos{(k_{z}c/2)}\\
 &  & +4t_{\alpha_{2},\alpha_{4}}^{010}e^{i(k_{y}a\frac{\sqrt{3}}{2})}\cos{(k_{x}a/2)}\cos{(k_{z}c/2)}\\
 &  & +4t_{\alpha_{2},\alpha_{4}}^{0\bar{1}0}e^{-i(k_{y}a\frac{\sqrt{3}}{2})}\cos{(k_{x}a/2)}\cos{(k_{z}c/2)}\\
H_{\alpha_{1},\alpha_{2}} & = & 2it_{\alpha_{1},\alpha_{2}}^{100}\sin{(k_{x}a)}+2it_{\alpha_{1},\alpha_{2}}^{010}e^{i(k_{y}a\frac{\sqrt{3}}{2})}\sin{(k_{x}a/2)}\\
 &  & +2it_{\alpha_{1},\alpha_{2}}^{0\bar{1}0}e^{-i(k_{y}a\frac{\sqrt{3}}{2})}\sin{(k_{x}a/2)}\\
H_{\alpha_{1},\alpha_{4}} & = & 4it_{\alpha_{1},\alpha_{4}}^{100}\sin{(k_{x}a)}\cos{(k_{z}c/2)}\\
 &  & +4it_{\alpha_{1},\alpha_{4}}^{010}e^{i(k_{y}a\frac{\sqrt{3}}{2})}\sin{(k_{x}a/2)}\cos{(k_{z}c/2)}\\
 &  & +4it_{\alpha_{1},\alpha_{4}}^{0\bar{1}0}e^{-i(k_{y}a\frac{\sqrt{3}}{2})}\sin{(k_{x}a/2)}\cos{(k_{z}c/2)}\\
H_{\alpha_{1},\alpha_{5}} & = & 4it_{\alpha_{1},\alpha_{5}}^{100}\sin{(k_{x}a)}\cos{(k_{z}c/2)}\\
 &  & +4it_{\alpha_{1},\alpha_{5}}^{010}e^{i(k_{y}a\frac{\sqrt{3}}{2})}\sin{(k_{x}a/2)}\cos{(k_{z}c/2)}\\
 &  & +4it_{\alpha_{1},\alpha_{5}}^{0\bar{1}0}e^{-i(k_{y}a\frac{\sqrt{3}}{2})}\sin{(k_{x}a/2)}\cos{(k_{z}c/2)}\\
H_{\alpha_{2},\alpha_{3}} & = & 4it_{\alpha_{2},\alpha_{3}}^{100}\sin{(k_{x}a)}\cos{(k_{z}c/2)}\\
 &  & +4it_{\alpha_{2},\alpha_{3}}^{010}e^{i(k_{y}a\frac{\sqrt{3}}{2})}\sin{(k_{x}a/2)}\cos{(k_{z}c/2)}\\
 &  & +4it_{\alpha_{2},\alpha_{3}}^{0\bar{1}0}e^{-i(k_{y}a\frac{\sqrt{3}}{2})}\sin{(k_{x}a/2)}\cos{(k_{z}c/2)}\\
H_{\alpha_{2},\alpha_{5}} & = & 4t_{\alpha_{2},\alpha_{5}}^{100}\cos{(k_{x}a)}\cos{(k_{z}c/2)}\\
 &  & +4t_{\alpha_{2},\alpha_{5}}^{010}e^{i(k_{y}a\frac{\sqrt{3}}{2})}\cos{(k_{x}a/2)}\cos{(k_{z}c/2)}\\
 &  & +4t_{\alpha_{2},\alpha_{5}}^{0\bar{1}0}e^{-i(k_{y}a\frac{\sqrt{3}}{2})}\cos{(k_{x}a/2)}\cos{(k_{z}c/2)}\\
H_{\alpha_{3},\alpha_{4}} & = & 2it_{\alpha_{3},\alpha_{4}}^{100}\sin{(k_{x}a)}+2it_{\alpha_{3},\alpha_{4}}^{010}e^{i(k_{y}a\frac{\sqrt{3}}{2})}\sin{(k_{x}a/2)}\\
 &  & +2it_{\alpha_{3},\alpha_{4}}^{0\bar{1}0}e^{-i(k_{y}a\frac{\sqrt{3}}{2})}\sin{(k_{x}a/2)}\\
H_{\alpha_{3},\alpha_{5}} & = & 2it_{\alpha_{3},\alpha_{5}}^{100}\sin{(k_{x}a)}+2it_{\alpha_{3},\alpha_{5}}^{010}e^{i(k_{y}a\frac{\sqrt{3}}{2})}\sin{(k_{x}a/2)}\\
 &  & +2it_{\alpha_{3},\alpha_{5}}^{0\bar{1}0}e^{-i(k_{y}a\frac{\sqrt{3}}{2})}\sin{(k_{x}a/2)}\\
H_{\alpha_{4},\alpha_{5}} & = & 2t_{\alpha_{4},\alpha_{5}}^{100}\cos{(k_{x}a)}+2t_{\alpha_{4},\alpha_{5}}^{010}e^{i(k_{y}a\frac{\sqrt{3}}{2})}\cos{(k_{x}a/2)}\\
 &  & +2t_{\alpha_{4},\alpha_{5}}^{0\bar{1}0}e^{-i(k_{y}a\frac{\sqrt{3}}{2})}\cos{(k_{x}a/2)}
\end{eqnarray*}

The numerical values of the hopping parameters in the above expressions
are reported in Tables \ref{TabNONDIAG1} and \ref{TabNONDIAG2}.
In particular we see from Table~\ref{TabNONDIAG2} that the most
relevant hopping amplitudes, larger than 30 meV, occur between the
planar $\alpha_{1}$ and $\alpha_{2}$ and between the planar $\alpha_{3}$
and $\alpha_{4}$ Wannier states in the $xy$ plane.

\begin{table}
\begin{centering}
\begin{tabular}{|c|c|c|c|c|c|c|}
\hline 
%{|p{16.0cm}|}
 & \multicolumn{3}{c|}{out of plane} & \multicolumn{3}{c|}{in plane}\tabularnewline
\hline 
 & 001 & 002 & 003 & 100 & 010 & $0\bar{1}0$\tabularnewline
\hline 
$\alpha_{1}\alpha_{3}$ & 7.9 & 7.3 & 14.6 & -7.4 & -6.9 & 1.4\tabularnewline
\hline 
$\alpha_{2}\alpha_{4}$ & 7.9 & 7.3 & 14.6 & -1.2 & -1.7 & -9.9\tabularnewline
\hline 
\end{tabular}
\par\end{centering}
\caption{Hopping integrals between $\alpha_{1}$-$\alpha_{3}$ and $\alpha_{2}$-$\alpha_{4}$
Wannier states (energy units in meV).}
\label{TabNONDIAG1}
\end{table}

\begin{table}
\begin{centering}
\begin{tabular}{|c|c|c|c|}
\hline 
%{|p{16.0cm}|}
 & 100 & 010 & $0\bar{1}0$\tabularnewline
\hline 
$\alpha_{1}\alpha_{2}$ & -15.0 & 30.3 & -0.2\tabularnewline
\hline 
$\alpha_{1}\alpha_{4}$ & -7.1 & -2.7 & 2.6\tabularnewline
\hline 
$\alpha_{1}\alpha_{5}$ & -9.1 & -8.6 & -0.5\tabularnewline
\hline 
$\alpha_{2}\alpha_{3}$ & -2.4 & -7.3 & -2.1\tabularnewline
\hline 
$\alpha_{2}\alpha_{5}$ & -4.7 & -5.6 & 10.2\tabularnewline
\hline 
$\alpha_{3}\alpha_{4}$ & 17.4 & 0.6 & -35.4\tabularnewline
\hline 
$\alpha_{3}\alpha_{5}$ & 13.0 & -1.0 & 14.1\tabularnewline
\hline 
$\alpha_{4}\alpha_{5}$ & -8.7 & 15.7 & -6.9\tabularnewline
\hline 
\end{tabular}
\par\end{centering}
\caption{Hopping integrals between different Wannier states, other than those
listed in Table II (energy units in meV).}
\label{TabNONDIAG2}
\end{table}
\newpage{}

\end{document}